\documentclass[4paper]{article}
\pdfoutput=1
\usepackage{jheppub}
\usepackage{epsfig,amsmath,latexsym,amssymb,centernot}
\usepackage{feynmp}

\renewcommand{\O}{{\cal O}}
 
\newcommand{\be}{\begin{equation}}
\newcommand{\ee}{\end{equation}}
\newcommand{\bea}{\begin{eqnarray}}
\newcommand{\ena}{\end{eqnarray}}

\newcommand{\ba}{\begin{eqnarray}}
\newcommand{\ea}{\end{eqnarray}}

\newcommand{\del}{\partial }

\title{Lectures on the Cosmological Constant Problem} 
\date{\today}

\author{Antonio Padilla}
\emailAdd{antonio.padilla@nottingham.ac.uk}
\affiliation{School of Physics and Astronomy, University of Nottingham, Nottingham NG7 2RD, UK}  

\date{\small \today}
\abstract{These lectures on the cosmological constant problem were prepared for the X Mexican School on Gravitation and Mathematical Physics. The problem itself is explained in detail, emphasising the importance of radiative instability and the need to {\it repeatedly} fine tune as we change our effective description. Weinberg's no go theorem is worked through in detail. I review a number of proposals including Linde's universe multiplication, Coleman's wormholes, the fat graviton, and SLED, to name a few. Large distance modifications of gravity are also discussed, with causality considerations pointing towards a global modification as being the most sensible option.  The global nature of the cosmological constant problem is  also emphasized, and as a result, the sequestering scenario is reviewed in some detail, demonstrating the cancellation of the Standard Model vacuum energy through a global modification of General Relativity. } 

\begin{document}
\maketitle

\section{Introduction} 
Anyone who is interested in theoretical physics and cosmology will have, at some point, thought about the cosmological constant problem.  Pauli is generally credited with being the first person to worry about the gravitational effects of zero point energies in the 1920s. He is said to have performed the relevant calculation in a cafe,  applying a cut-off  in zero point energies at the classical electron radius then noting with amusement that the radius of the world "nicht einmal bis zum Mond reichen w\"urde" [would not even reach to the Moon] \cite{enz}. It was Zel'dovich who first published a serious discussion of the problem in a pair of papers in the late sixties \cite{Zel}. In the 50 years since then the cosmological constant problem has remained ``the mother of all physics problems"\footnote{This quote is usually attributed to Susskind.}.

Before describing the cosmological constant problem in greater detail in the next section, let me highlight those reviews I have found most useful in building my understanding of it. There is, of course, Weinberg's classic \cite{Wein}, which is essential reading for anyone interested in this topic. More recent reviews by Polchinski \cite{Pol} and Burgess \cite{Cliff} both provide an excellent discussion of the problem with useful insights, whilst going on to describe particular proposals in greater detail (respectively the landscape, and SLED). Martin's review \cite{Martin} is nice in that it contains all those nasty QFT calculations explicitly presented, while Nobbenhuis \cite{Nobb} provides a  comprehensive collection of credible attempts to tackle the problem. My notes are intended to compliment all of the above, and at the same time provide further insights inspired by countless discussions with my long term collaborator, Nemanja Kaloper, and the subsequent development of the sequestering proposal \cite{KP1,KP2,KP3}. I will focus more on (a subset of) theoretical ideas rather than a detailed discussion of observational data (for that, see e.g.  \cite{ed,Carroll1,Carroll2})

In some respects, the biggest problem with the cosmological constant problem is that it is rarely stated properly. Why is the cosmological constant so small? Why is it not Planckian? Why is it not zero? All of these questions belittle the cosmological constant problem, and if we are to have any hope of solving it, we had better be clear about what the problem really is.  As I will emphasise in these notes, the real issue with the cosmological constant is not so much one of fine tuning, but of radiative instability, and the need to {\it repeatedly} fine tune whenever the higher loop corrections are included.  This  reflects the cosmological constant's sensitivity to the details of the (unknown) UV physics and challenges our faith in  effective field theory - an essential tool for doing physics on large scales without detailed knowledge of the microphysics and unlimited calculational power.  Of course, it may be that there are a whole landscape of possible effective theories, and ours is selected not by naturalness, but by anthropic considerations\footnote{For an introduction to the cosmological constant problem, and the landscape, I recommend the reviews by Polchinski \cite{Pol} and Bousso \cite{Bousso}.}.

What we would really like to do is to render the cosmological constant radiatively stable somehow.  That is not to say we expect to predict its value. On the contrary, standard renormalisation methods suggest it will depend on an arbitrary subtraction scale and should therefore be {\it measured}, just as we are required to measure the electron charge.    We know from observation that the renormalised cosmological constant should not exceed the dark energy scale $(\text{meV})^4$.  However, the point about a {\it radiatively stable} cosmological constant is that its 
 measured value is robust against any change in the effective description. Of course, achieving this goal is a non-trivial task, and I will use these lectures to discuss many of the most interesting and innovative ideas that have been proposed. This will range from symmetry proposals to modifications of gravity but will {\it not} include any anthropics.

These notes are organised as follows. The next section will be devoted to a proper description of the cosmological constant problem in terms of radiative instability of the vacuum curvature. Section \ref{sec:not} will include some discussion of how {\it not} to solve the problem. This includes a review of Weinberg's famous no-go theorem \cite{Wein} forbidding so-called self adjustment mechanisms under certain key assumptions. I will also take the opportunity to repeat why unimodular gravity \cite{finkel,buch,henteitel,ippo} is unable to help with the cosmological constant problem. Section \ref{sec:symm} is devoted to symmetry and a discussion of technical naturalness. I will touch on supersymmetry, scale invariance and recall Linde's ingenious  universe multiplication proposal\cite{linde2} which exhibits an energy parity symmetry. Coleman's old wormhole idea \cite{cole} will be discussed in section \ref{sec:cole}, and   modified gravity proposals   in sections \ref{sec:short} and \ref{sec:long}, (corresponding to short and long distance modifications respectively).  Relevant short distance modifications of gravity include the fat graviton \cite{fat} and supersymmetric large extra dimensions (SLED) \cite{SLED}. Long distance modifications of gravity have been particularly popular recently, and I will advocate the ultimate long distance i.e. {\it a global} modification of gravity. This is realised by the sequestering scenario \cite{KP1,KP2,KP3} which will be discussed in some detail. I offer some final thoughts in section \ref{sec:summary}, paying particular attention to the robustness of the General Relativity.

\section{What is the problem?} \label{sec:ccp}
The cosmological constant problem is the unwanted child of  two pillars of twentieth century physics: quantum field theory and general relativity.  Thanks to locality and unitarity, quantum field theory tells us that the vacuum has an energy.  The grown up way to calculate this is to compute the vacuum loop diagrams for each particle species.  For example, the one loop diagram for a canonical scalar, $\phi$, of mass, $m$ yields, 
\begin{eqnarray}  
\unitlength=1mm
\begin{fmffile}{1loopvac}
\parbox{15mm}{
\begin{fmfgraph*}(15,15)
\fmfi{plain}{fullcircle scaled .5w shifted (.5w,.5h)}
\end{fmfgraph*} }
\end{fmffile} \sim ~\frac{i}{2} \text{tr}\left[\log \left(-i \frac{\delta^2 S}{\delta \phi(x) \delta \phi(y)} \right)\right]&=& -\frac12 \int d^4 x \int \frac{d^4 k_E}{(2\pi)^4} \log(k_E^2+m^2) \nonumber \\
&=&- \frac{m^4}{(8 \pi)^2} \left[ -\frac{2}{\epsilon} +\log \left(\frac{m^2}{4\pi \mu^2} \right) +\gamma-\frac32 \right] \int d^4 x \nonumber \\
&\subset& -V_{vac} \int d^4 x 
\end{eqnarray}
where $\gamma$ is the Euler-Mascheroni constant, and we have used dimensional regularisation to perform the momentum integral, having introduced an arbitrary mass scale, $\mu$. If we add counterterms to eliminate the divergences (we will come back to this point later), then we might expect that the total vacuum energy is given by
\be \label{m^4}
V_{vac} \sim \sum_{particles} {\cal O}(1) m_{particle}^4 
\ee
Given our knowledge of the Standard Model, which already includes particles more or less all the way up to the TeV scale, we would estimate $V_{vac}  \gtrsim (\text{TeV})^4$. Of course, in the absence of gravity these zero point contributions do not affect the dynamics. However, when (classical) gravity is switched on the equivalence principle tells us that all forms of energy curve spacetime, and the vacuum energy is no exception. General coordinate invariance then tells us  {\it how} vacuum energy gravitates, as we introduce a covariant measure
\be \label{covcoupling}
-V_{vac} \int d^4 x  \longrightarrow -V_{vac} \int d^4 x \sqrt{-g} 
\ee
where $g=\det g_{\mu\nu}$. In terms of Feynman diagrams we can think of this as coming from the sum of vacuum loops connected to external graviton fields, i.e. 
%\be
%\begin{fmffile}{1loopvac1}
%  \begin{fmfgraph}(40,25)
%    \fmfleft{i}
%    \fmfright{o}
%    \fmf{photon}{i,v1}
%    \fmf{plain,left}{v1,o,v1}
%  \end{fmfgraph}
%  \end{fmffile}
%\ee

\be 
\unitlength=1mm
\begin{fmffile}{1loopvac0}
\parbox{15mm}{
 \begin{fmfgraph}(15,15)
    \fmfleft{i,i1,i2}
    \fmfright{o,o1,o2}
    \fmf{phantom}{i2,v1}
    \fmf{plain,left=0.4}{v1,v2,v3,v4,v1}
    \fmf{phantom}{v2,o2}
     \fmf{phantom}{i,v4}
      \fmf{phantom}{v3,o}
  \end{fmfgraph}}
\end{fmffile} \qquad +
\unitlength=1mm
\begin{fmffile}{1loopvac1}
\parbox{15mm}{
 \begin{fmfgraph}(15,15)
    \fmfleft{i,i1,i2}
    \fmfright{o,o1,o2}
    \fmf{photon}{i2,v1}
    \fmf{plain,left=0.4}{v1,v2,v3,v4,v1}
    \fmf{phantom}{v2,o2}
     \fmf{phantom}{i,v4}
      \fmf{phantom}{v3,o}
  \end{fmfgraph}}
\end{fmffile} \qquad +
\unitlength=1mm
\begin{fmffile}{1loopvac2}
\parbox{15mm}{
 \begin{fmfgraph}(15,15)
    \fmfleft{i,i1,i2}
    \fmfright{o,o1,o2}
    \fmf{photon}{i2,v1}
    \fmf{plain,left=0.4}{v1,v2,v3,v4,v1}
    \fmf{photon}{v2,o2}
     \fmf{phantom}{i,v4}
      \fmf{phantom}{v3,o}
  \end{fmfgraph}}
\end{fmffile}\qquad + \quad \ldots \quad 
 \sim \quad   -V_{vac} \int d^4 x \sqrt{-g} 
 \ee
\subsection{Do vacuum fluctuations really exist?} \label{sec:exist}

\begin{figure}[h]
\begin{center}
\epsfig{file=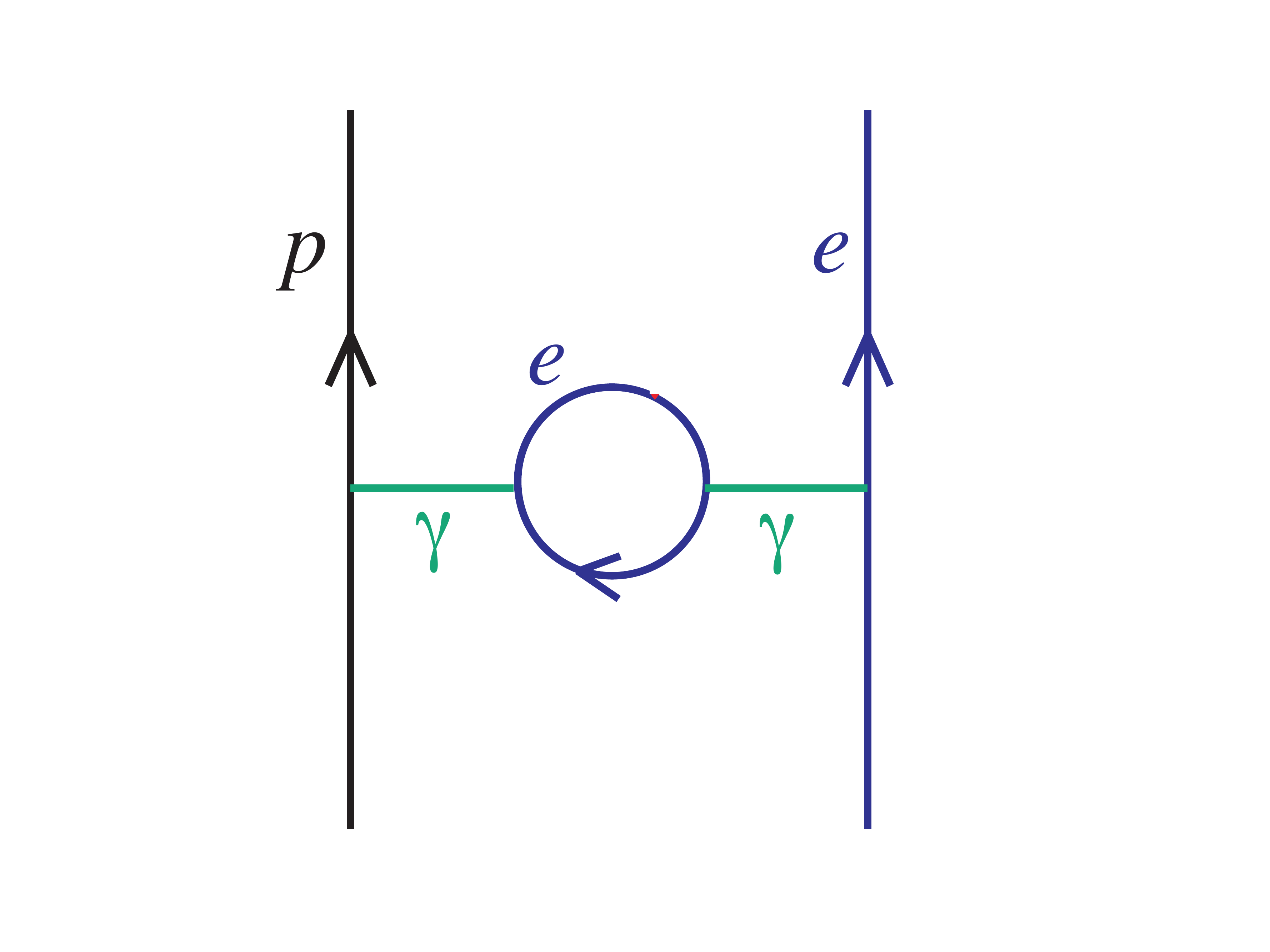, width= 70 mm, height= 50 mm}
\epsfig{file=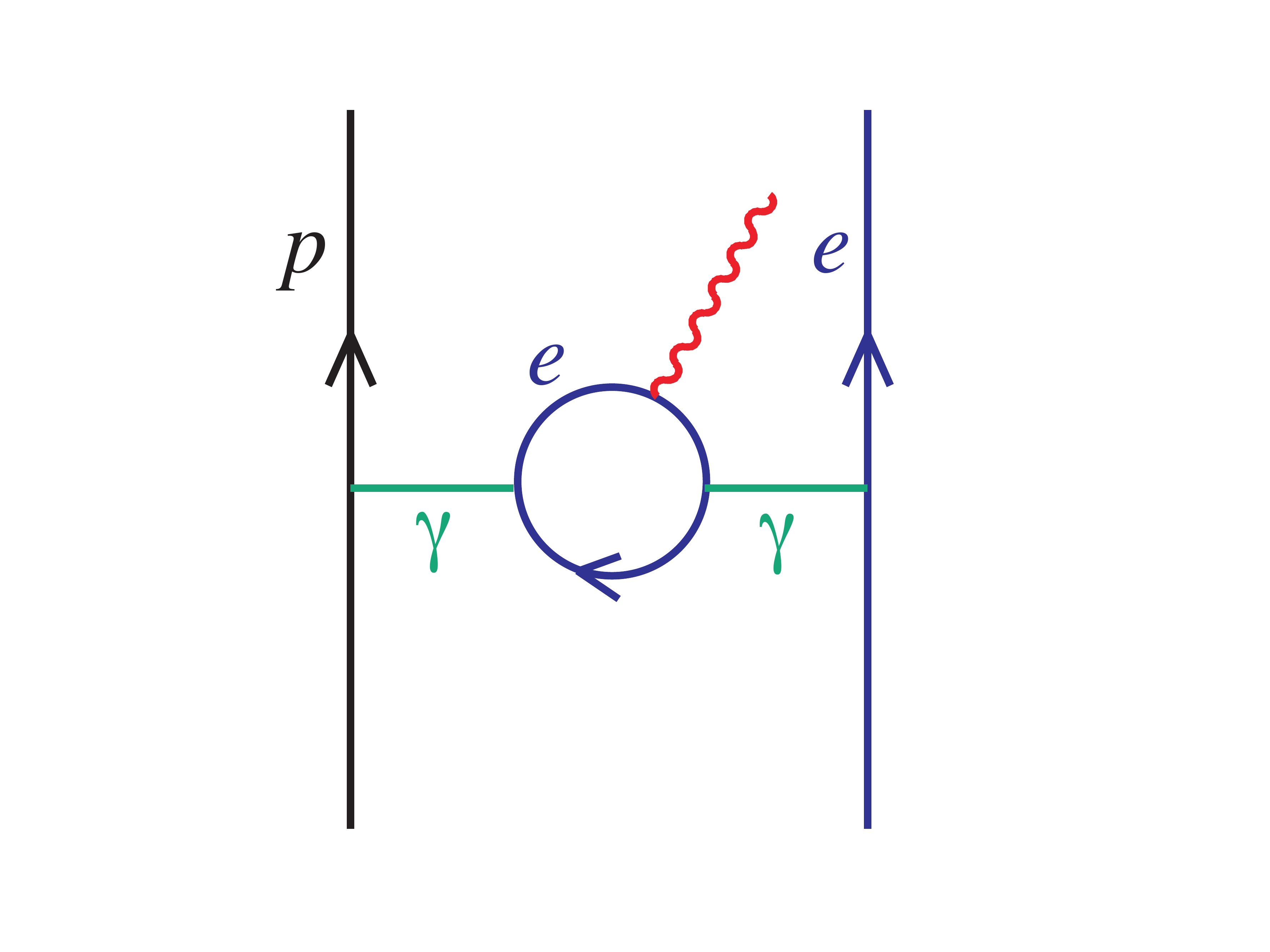, width= 70 mm, height= 50 mm}
\caption{Figure adapted from \cite{Pol}, showing the one loop vacuum polarization  contribution to the Lamb shift, with and without coupling to an external graviton (wiggly red line).} \label{fig:lamb}
\end{center}
\end{figure}

 The answer would seem to be that they do exist, as evidenced by the Lamb shift \cite{IZ} or the Casimir effect \cite{cas}\footnote{However, as Jaffe has emphasised  \cite{jaffe}, the calculation of both of these effects correspond to Feynman diagrams with externals legs of Standard Model fields. A true vacuum bubble has no external legs at all, and from a gravitational perspective the diagrams we are interested in only have external graviton lines.}. For example, for the Lamb shift, recall that vacuum polarisation contributions  like those shown  to the left in Figure \ref{fig:lamb} imply that the energy levels of the Hydrogen atom are shifted by an amount (see \cite{Pol,Martin} for further details)
\be
\Delta E(n, l) \approx \delta_{0 l} \frac{4 m_e}{3 \pi n^3} \ln (1/\alpha) Z
\ee
where $n$ describes the energy quantum number, $l=0, \ldots n-1$ the angular momentum quantum number, and $Z$ the atomic number ($Z=1$ for Hydrogen), so that degenerate states with $n=2$ are split according to angular momenta.
But even if vacuum fluctuations exist, can we be sure that they gravitate? Again, it would seem so because the Lamb shift  not only affects the inertial energy of an atom, but also its gravitational energy. Indeed, for heavy nuclei such as Aluminium and Platinum  we can compare the ratio of gravitational mass to inertial mass for each \cite{EP}.  It turns out that the ratios stay the same (to order $10^{-12}$) even though the effect of vacuum polarisation differs by a factor of $3$ for the inertial masses $\left(\frac{\Delta E_{Al}}{E_{Al}} \sim 10^{-3}, \frac{\Delta E_{Pt}}{E_{Pt}} \sim 3. 10^{-3}\right)$. 

\subsection{How does vacuum energy affect spacetime?}   At low enough energies (below TeV), let us assume that Nature admits an effective description in terms of quantum matter  coupled to classical gravity.  This semi-classical picture is certainly valid if gravity is described by General Relativity up to the Planck scale, so that we have Einstein's equation
\be
M_{pl}^2 G_{\mu\nu}=T_{\mu\nu}
\ee
relating the classical geometry of spacetime to the expectation value of the energy momentum tensor for the  quantum matter fields. The covariant coupling  (\ref{covcoupling}) suggests that vacuum energy contributes an energy momentum tensor 
\be \label{Tvac}
T^{vac}_{\mu\nu}=-V_{vac} g_{\mu\nu}
\ee
As they correspond to (infinitely) long wavelength modes, the effect of vacuum energy is most pronounced on cosmology. Neglecting any excitations about the vacuum at late times, we see that the vacuum curvature is given by $H_{vac}^2 =\frac{V_{vac}}{3M_{pl}^2} $. If $V_{vac}>0$ this yields an accelerated late time de Sitter expansion which could be compatible with current observation provided $H_{vac}^2 \lesssim H_0^2 \sim (\text{meV})^4/M_{pl}^2$. Note that if the vacuum curvature exceeded this bound it would have  started to dominate long before the current epoc.

Now, recall from (\ref{m^4}) that $V_{vac}$ receives contributions from each particle species that go like the fourth power of the mass.  Focussing only on (say) the contribution of the electron, so that $V_{vac} \sim m_e^4$, and applying Einstein's equations, we see that the cosmological horizon lies at a distance $r_H \lesssim 1/H_{vac}$, where $H_{vac}^2 \sim m_e^4/M_{pl}^2$. This yields
\be
r_H \lesssim 10^6 \text{km} \sim \text{Earth-Moon distance}
\ee 
reminding us of Pauli's colourful quote \cite{enz}.  Of course, current observations tell us that $r_H \sim 1/H_0 \sim 10^{26}$m  so the electron already appears to provide way too much vacuum energy. Heavier particles only make the situation worse, and we know these exist  at least up the electroweak scale and most probably well beyond. 

How do we reconcile this? Well, we recall that we have already introduced a counterterm to absorb the divergences  in the vacuum energy. Let us include this explictly, so that our action is given by
\be \label{GRact}
S=\int d^4 x \sqrt{-g} \left[ \frac{M_{pl}^2}{2}R -{\cal L}_m(g^{\mu\nu}, \Psi) - \Lambda \right]
\ee
where ${\cal L}_m$ is the effective Lagrangian for the matter fields, $\Psi$. This contributes a divergent vacuum energy. The ``counterterm" $\Lambda$ is also divergent, so that what actually gravitates is the finite combination
\be
\Lambda_{ren}=\Lambda+V_{vac}
\ee
Observations now require $\Lambda_{ren} \lesssim (\text{meV})^4$   so {\it  ignoring the divergences} this represents a fine tuning of $\lesssim 10^{-60}$ between the finite parts of $\Lambda$ and $V_{vac}$ since
\be
V_{vac}^{finite} \gtrsim (\text{TeV})^4 \sim 10^{60} (\text{meV})^4
\ee
 This is often what people call the cosmological constant problem, but of course this cannot be anything like the full story! We have happily cancelled divergences, so why are we getting upset about cancelling a couple of big (but not infinite) numbers? There must be more to it.

\subsection{The real problem: radiative instability}
 To describe the quantum matter sector, let us do the one thing we know how to do: perturbation theory.   We shall illustrate the radiative instability of the vacuum energy by again focussing on a single scalar of mass, $m$,  with $\lambda \phi^4$ self coupling, minimally coupled to the classical graviton.  The results are qualitatively the same for fermions and gauge bosons.  In any event, the one loop contribution to the vacuum energy is given by
\be
V_{vac}^{\phi, 1loop} \sim  - \frac{m^4}{(8 \pi)^2} \left[ \frac{2}{\epsilon} +\log \left(\frac{\mu^2}{m^2} \right) +\text{finite} \right] 
\ee
The divergence requires us to add the following counterterm which depends on an {\it arbitrary subtraction scale}, $M$, 
\be
\Lambda^{1loop}  \sim   \frac{m^4}{(8 \pi)^2} \left[ \frac{2}{\epsilon} +\log \left(\frac{\mu^2}{M^2} \right)  \right] 
\ee
so that the renormalised vacuum energy (at one loop) is given  by
\be
\Lambda^{1loop}_{ren}\sim \frac{m^4}{(8 \pi)^2} \left[ \log \left(\frac{m^2}{M^2} \right) -\text{finite} \right] 
\ee
Because this depends explicitly on the arbitrary scale $M$, we do not have a concrete prediction for the renormalised vacuum energy, so it doesn't really make much sense at this stage to say what the answer should or should not be. Instead, what we need to do is go out and measure $\Lambda_{ren}$.  As we have already seen, with heavy particles contributing more or less up to the TeV scale (and possibly beyond) such a measurement suggests the finite contributions to the one loop renormalised vacuum energy are cancelling to at least an accuracy of one part in $10^{60}$ in Nature. 

At this stage of the argument, we have no issue with the above fine tuning. The issue arises when we alter our effective description for matter by going to two loops. For the toy scalar described above we consider the so-called ``figure of eight" with external graviton legs.  This yields a two loop correction to the vacuum energy that scales as $\lambda m^4$.  For perturbative theories without finely tuned couplings this means the two loop correction is not significantly suppressed with respect to the one loop contribution\footnote{For example, for the Standard Model Higgs we have $\lambda \sim {\cal O}(0.1)$}. Therefore, the cancellation we imposed at one loop is completely spoilt, and we must {\it retune} the finite contributions in the counterterm to more or less the same degree of accuracy.

Now if we have particularly thick skin, we might even accept this repeated fine tuning at both one and two loops. But then we go to three loops, then four, and so on. At each successive order in perturbation theory we are required to fine tune to extreme accuracy.  This is radiative instability.  There is no point in tuning, say, the one loop counter term, because that tuning will always be unstable to higher loops. What this is telling us is that the vacuum energy is uber-sensitive to the details of UV physics of which we are ignorant, the sensitivity encoded in the loop instability of the effective field theory. This is the cosmological constant problem.

One might argue that all of the above is really just an artefact of perturbation theory, and that in reality one should sum all the loops to obtain an exact expression for the vacuum energy.  Thats fine, but for which theory should we do this? The fact is that the full loop structure depends on the full effective field theory expansion, way beyond leading order, with all the unknown coefficients reflecting the unknown UV physics. Whilst there is only one dimension five operator correcting the leading order Standard Model EFT, there are $59$ operators of dimension six \cite{59}, and untold number beyond that. It is optimistic,  to say the least, to suggest we have any hope of ``summing all the loops".

\subsection{An effective description of the problem}  There is an alternative, although entirely equivalent, way to describe the cosmological constant problem that does not rely on changing the loop order in perturbation  theory.  This uses the Wilson action, where it turns out that the vacuum energy is unstable against changing the Wilsonian cut-off. This is described very elegantly in \cite{Cliff} but we repeat some of the main points here for completeness.  The basic idea of the Wilson action is to integrate out heavy degrees of freedom above some mass scale $\mu$. To this end we split the degree of freedom up as $\phi=\phi_l+\phi_h$, with the light modes, $\phi_l$, having energies  less than $\mu$, and the heavy modes $\phi_h$, having energies greater than $\mu$. The Wilson effective action, $S_\text{eff}[\phi_l]$ is given by integrating out the heavy modes in the path integral
\be
\exp(i S_\text{eff}[\phi_l])=\int D \phi_h \exp(i S[\phi_l, \phi_h])
\ee
where $S[\phi_l, \phi_h]$ is the full microscopic action including both light and heavy degrees of freedom. A computation of the vacuum energy using the effective action $S_\text{eff}[\phi_l]$ valid up to the scale $\mu$, would be expected to be of order $\mu^4$. This fixes the cosmological counterterm that we add in order match the renormalised cosmological constant to the low energy observation.

However, consider what happens when we move the cut-off to a new scale $\hat \mu <\mu$. This involves further  integrating out those degrees  of freedom with energies, $E$, lying in the interval $\hat \mu<E< \mu$. The resulting effective action $\hat S_\text{eff}$ yields a vacuum energy that goes as $\hat \mu^4$ which ought to be cancelled by the cosmological counterterm.  But the cosmological counterterm was already fixed to guarantee cancellation in the original  effective field theory, with the higher cut-off $\mu$. We have to  retune it.  This isn't how effective actions are supposed to work -- we should not have to choose a completely different low energy counterterm simply because we integrated out a few more high energy modes.   Once again  we are seeing uber-sensitivity to the details of the theory in the UV. 

The bottom line is that the tuning of the cosmological constant required to match observation is unstable against any change in the effective description, be it changing the loop order in perturbation theory, or changing the position of the Wilsonian cut-off.   It is completely analogous to the problem facing the Higgs mass, although there the tunings are less severe and supersymmetry provides an elegant solution.

\subsection{A further complication: phase transitions}  Phase transitions in the early Universe can cause the vacuum energy to jump by a finite amount over a period of time \cite{PT}. Therefore, if we managed to achieve the desired cancellation of the cosmological constant before the transition, it would be spoilt afterwards (or vice versa).  Note that the electroweak phase transition would be expected to yield a jump of order $\Delta V_{EW} \sim (200 ~\text{GeV})^4$, whilst the QCD transition yields a jump $\Delta V_{QCD} \sim (0.3  ~\text{GeV})^4$. 

\subsection{Two big questions} If we are to solve the cosmological constant problem, we must come up with a way to make the observed cosmological constant stable against radiative corrections. We are happy to tolerate one off fine tunings, but not order by order {\it re}tunings.  It is important to note that within the context of effective field theory and  renormalisation we should not expect a radiatively stable cosmological constant to be predictable. Like the coefficient of any operator with dimension $\leq 4$, it depends on an arbitrary subtraction scale and should therefore be measured. This is OK as long as nothing drastic happens when we tweak our effective description e.g. by adding higher loops or by moving the Wilsonian cut-off.

This discussion leaves us with two very important questions: (1) how do we really measure $\Lambda$? and (2) how do we make $\Lambda$ radiatively stable. In section \ref{sec:long} we will argue that the answer to both these questions is the same: go global!

\section{Some things not to do} \label{sec:not}
The main purpose of this section is to review Weinberg's famous no go theorem \cite {Wein} which acts like a straight jacket, prohibiting  many solutions to the cosmological constant problem. Indeed,  the first question any theorist should ask when he/she thinks they might have  a solution is: how do I get around Weinberg no go? Of course, whilst a straight jacket restricts us in some ways, it doesn't restrict us in every way, and there is a clear sense in which Weinberg's theorem should  point us in other, more promising, directions by relaxing his assumptions.  

We will  also discuss unimodular gravity \cite{finkel,buch,henteitel,ippo}, which is another no go area when it comes to the cosmological constant problem.  In fact, unimodular gravity brings no new perspective to the problem whatsoever and understanding why is something of a litmus test in terms of understanding the problem itself. 

\subsection{Unimodular gravity} The basic idea behind unimodular gravity is to perform a restricted variation of the Einstein-Hilbert action, assuming $|\det g|=1$.  This might seem like a clever thing to do since the vacuum energy couples to something which is not allowed to vary, and so one might expect it to drop out of the dynamics. The trouble is that, locally at least, $|\det g|=1$ is merely a gauge choice in General Relativity, and since gauge symmetry represents nothing more than a redundancy of description, it is difficult to see how this could help in any way.  

The field equations that arise from the restricted variation correspond to the traceless Einstein equations
\be \label{umgeq}
M_{pl}^2 \left[ R_{\mu\nu}-\frac14 R g_{\mu\nu} \right]=T_{\mu\nu}-\frac14 T g_{\mu\nu}
\ee
The vacuum contribution to the energy momentum tensor (\ref{Tvac}) is a pure trace, so one might happily conclude that the vacuum energy does indeed drop out. Not so fast!  If we take the divergence of equation (\ref{umgeq}), making use of both the Bianchi identity ($\nabla^\mu G_{\mu\nu}=0$) and energy conservation ($\nabla^\mu T_{\mu\nu}=0$) we find 
\be
\nabla_\mu (T+M_{pl}^2 R)=0 \implies T+M_{pl}^2 R=4 \Lambda
\ee
where $\Lambda$ is a constant of integration. Plugging this back into the unimodular equation of motion (\ref{umgeq}) we get
\be
M_{pl}^2 G_{\mu\nu}=T_{\mu\nu}-\Lambda g_{\mu\nu}
\ee
Thus the integration constant $\Lambda$ is playing the same role as the ``counterterm" discussed in the context of General Relativity.  This means it can eat the divergences in $V_{vac}$, but there is no way it can cope  with additional loop corrections. In other words, the integration constant must be retuned order by order in perturbation theory just like the counterterm in GR. We have gained nothing. 

At first glance, this conclusion might seem puzzling.  By fixing $|\det g|=1$ it looks as if  the cosmological constant is not a coupling of a dynamical operator, so no quantum fluctuations of any field should affect its value. However, it is important to realise that the restriction on $\det g$ should be properly implemented and one way to do this is via  a Lagrange multiplier, so that the action for unimodular gravity can be expressed as
\be
S_\text{UMG}=\int d^4 x \left[  \frac{M_{pl}^2}{2} \sqrt{-g} R - \sqrt{-g}  {\cal L}_m(g^{\mu\nu}, \Psi) +\lambda (\sqrt{-g}-1)  \right]
\ee
Radiative corrections to the vacuum energy now  shift the Lagrange multiplier, $\lambda$, rendering its boundary value radiatively unstable. However, the boundary value of $\lambda$ is precisely the integration constant, $\Lambda$,  discussed above, and is equivalent to the cosmological constant in GR.

Another way to implement the constraint $|\det g|=1$ is as in Weyl-Transverse Gravity \cite{wtg} (which is a special case of \cite{ian}), where the  constrained metric, $g_{\mu\nu}$, entering the Einstein-Hilbert action is written in terms of an unconstrained metric, $f_{\mu\nu}$, where $g_{\mu\nu}=\frac{f_{\mu\nu}}{|\det f|^{1/4}}$. The renormalised cosmological constant   now enters the action  as $(\text{constant}) \times  \int d^4 x$, which one might erroneously interpret as non-dynamical. However, by a simple change of coordinates we see that this does give dynamics because, in the absence of full diffeomorphism invariance, it depends explicitly on a dynamical Jacobian\footnote{Let $x^\mu \to X^\mu(x)$, then $\int d^4 x \to \int d^4 x \left| \frac{\del X}{\del x} \right|$, and $\frac{\delta }{\delta X^\mu}  \int d^4 x \left| \frac{\del X}{\del x} \right| \neq 0$ (see  \cite{kuchar}). }. For a more detailed discussion of the issues relating the cosmological constant problem to unimodular gravity, we refer the reader to \cite{ippo}.

\subsection{Weinberg's no go theorem} The basic idea of {\it  self-adjustment } or {\it self tuning} is to add extra fields to the matter sector whose job is to ``eat up" the large vacuum energy, protecting the spacetime curvature accordingly. Weinberg's venerable no go theorem argues, under very general assumptions, that this is not possible without fine tuning.  Let us repeat his argument.

We begin by assuming the following field content: a spacetime metric, $g_{\mu\nu}$, and self adjusting matter fields, $\varphi_i$, with the tensor structure suppressed. The dynamics is described by a general Lagrangian density ${\cal L}[g, \varphi_i]$. We further assume that the vacuum is {\it translationally invariant}, so that on-shell we have $g_{\mu\nu}, \varphi_i=$ constant. This leaves a residual GL(4) symmetry,  given by a coordinate change $x^\mu \to (M^{-1})^\mu{}_\nu x^\nu$, where $M^\mu{}_\nu $ is a constant $4 \times 4 $ matrix. Note that the metric transforms as
\be
g_{\mu\nu} \to g_{\alpha \beta}M^\alpha{}_\mu M^\beta{}_\nu
\ee
and the Lagrangian density as
\be
{\cal L}(g, \varphi_i) \to \det M {\cal L}(g, \varphi_i)
\ee
We can write these transformations infinitesimally as
\be \label{GL4}
\delta_{\delta M} g_{\mu\nu}=\delta M_{\mu\nu}+\delta M_{\nu\mu}, \qquad \delta_{\delta M} {\cal L}=\text{Tr} \delta M {\cal L} 
\ee
 and because we have a constant vacuum solution, we can also infer the following relation
\be 
 \delta_{\delta M} {\cal L}=\frac{\del {\cal L}}{\del \varphi_i}  \delta_{\delta M} \varphi_i +\frac{\del {\cal L}}{\del g_{\mu\nu}}  \delta_{\delta M}  g_{\mu\nu}
 \ee
Meanwhile, the vacuum field equations are given by
\be \label{vaceqs}
\frac{\del {\cal L}}{\del \varphi_i}=0, \qquad 
 \frac{\del {\cal L}}{\del g_{\mu\nu}}=0 
\ee
 We now consider two distinct scenarios: (1) where both equations in (\ref{vaceqs}) hold independently, and (2) where they do not hold independently.  For the first scenario, we may assume $\frac{\del {\cal L}}{\del \varphi_i}=0$ without, for the moment,  assuming $\frac{\del {\cal L}}{\del g_{\mu\nu}}=0 $. Then our expressions for $ \delta_{\delta M} {\cal L}$ imply a relation
 \be
 \frac{\del {\cal L}}{\del g_{\mu\nu}} (\delta M_{\mu\nu}+\delta M_{\nu\mu})=\text{Tr} \delta M {\cal L}  \implies  \frac{\del {\cal L}}{\del g_{\mu\nu}}=\frac12 g^{\mu\nu} {\cal L}
 \ee
 which we can solve explicitly to give
 \be
 {\cal L}=\sqrt{-g} V(\varphi_i)
 \ee
The remaining field equation $\frac{\del {\cal L}}{\del g_{\mu\nu}}=0 $ now yields $V(\varphi_i)=0$, which corresponds to fine tuning.

We now consider the second scenario assuming that the equations in (\ref{vaceqs})  do not hold independently thanks to the following relation (which Weinberg assumes)
\be \label{relation}
2 g_{\mu\nu} \frac{\del {\cal L}}{\del g_{\mu\nu}}= \sum_i f_i(\varphi) \frac{\del {\cal L}}{\del \varphi_i}
\ee
where the $f_i$ depend on the self-adjusting fields. This relation represents a degree of degeneracy which we might well have expected given that we would like self-adjustment to occur for any vacuum energy. It also suggests a new scaling symmetry
\be
\delta_\epsilon g_{\mu\nu}=2 \epsilon g_{\mu\nu}, ~ \delta_\epsilon \varphi_i=-\epsilon f_i 
\ee
since then $\delta_\epsilon{\cal  L}=0$. We can transform\footnote{Think of the transformation $\delta_\epsilon \varphi_i$ as being generated by $X= \sum_i f_i(\varphi) \frac{\del ~}{\del \varphi_i} =\frac{\del ~}{\del \tilde  \varphi_0}$. Then the $\frac{\del ~~}{\del \tilde  \varphi_{i\neq 0}}$ represent tangents lying in the plane orthogonal to $X$. } the $\varphi_i$ in field space, $\varphi_i \to \tilde \varphi_i$ such that the scaling relation can be written as
\be
\delta_\epsilon g_{\mu\nu}=2 \epsilon g_{\mu\nu}, ~ \delta \tilde \varphi_0=-\epsilon, ~\delta \tilde \varphi_{i \neq 0}=0
\ee
 and because $\delta_\epsilon \left( e^{2\tilde \varphi_0} g_{\mu\nu}\right)=0$, we conclude that
\be
{\cal L}={\cal L} \left( e^{2\tilde \varphi_0} g_{\mu\nu}, \tilde \varphi_{i \neq 0}\right)
\ee
We now return to the GL(4) symmetry, which gives the transformations (\ref{GL4}) along with the relation 
\be 
 \delta_{\delta M} {\cal L}=\frac{\del {\cal L}}{\del \tilde \varphi_0}  \delta_{\delta M} \tilde \varphi_0+\frac{\del {\cal L}}{\del \tilde \varphi_{i \neq 0}}  \delta_{\delta M} \tilde \varphi_{i\neq 0} +\frac{\del {\cal L}}{\del g_{\mu\nu}}  \delta_{\delta M}  g_{\mu\nu}
 \ee
Note also that $\tilde \varphi_0$ is a scalar  and so it does not transform, $\delta_{\delta M} \tilde \varphi_0=0$. Having accounted for the degeneracy, we now have the following {\it independent} field equations on the vacuum
\be \label{vaceqs1}
\frac{\del {\cal L}}{\del \tilde \varphi_{i \neq 0}}=0, \qquad 
 \frac{\del {\cal L}}{\del g_{\mu\nu}}=0 
\ee
Proceeding exactly as we did earlier,  we assume $\frac{\del {\cal L}}{\del \tilde \varphi_{i \neq 0}}=0$ holds without  assuming $ \frac{\del {\cal L}}{\del g_{\mu\nu}}$ then use our expressions for $\delta_{\delta M} {\cal L}$ to derive the solution
\be
{\cal L}=\sqrt{-\det (e^{2\tilde \varphi_0} g_{\mu\nu})} V ( \tilde \varphi_{i \neq 0})=\sqrt{-g} e^{4 \tilde \varphi_0}V(  \tilde \varphi_{i \neq 0})
\ee
The remaining field equation $\frac{\del {\cal L}}{\del g_{\mu\nu}}=0 $ now yields $e^{4 \tilde \varphi_0}V(\varphi_{i \neq 0})=0$, which is subtlely different to what we had earlier. One possibility is that $V(\varphi_{i \neq 0})=0$, which represents fine tuning, while the other possibility is that $e^{\tilde \varphi_0} \to 0$.  Now since $e^{2\tilde \varphi_0} $ always comes with $g_{\mu\nu}$ via a conformal factor, it follows that all masses scale as $e^{\tilde \varphi_0} $.  Therefore, the limit $e^{\tilde \varphi_0} \to 0$ corresponds to a scale invariant limit where all masses go to zero, and that is simply not our Universe.

 This completes Weinberg's theorem and any proposal for solving the cosmological constant problem must explain how it escapes its clutches. This is especially true when we consider the fact that on small enough scales one will generically be able to assume approximate translation invariance and constancy of the fields.

\section{Symmetry}  \label{sec:symm}
Whenever a parameter in our theory vanishes, or is anomalously small, it is tempting to think a symmetry is at work.  The classic example of this is the electron mass, which lies many orders of magnitude below the electroweak scale. Why is the electron so light? The reason is that chiral symmetry kicks in as the electron mass goes to zero \cite{thooft}. Is there a symmetry protecting the cosmological constant?

\subsection{'tHooft naturalness} The interplay between the electron mass and chiral symmetry is an example of {\it 'tHooft naturalness}.  This states that a parameter $\epsilon$ is naturally small if there exists an enhanced symmetry in the limit $\epsilon \to 0$.  To see why this makes sense consider a Lagrangian ${\cal L}_0[\varphi]$ obeying some global symmetry $\phi \to \phi'$, that remains valid at the quantum level. If we add an operator $O[\varphi]$ that breaks the symmetry then the Lagrangian
\be
{\cal L}_\epsilon[\varphi]={\cal L}_0[\varphi]+\epsilon O[\varphi]
\ee
contains a parameter $\epsilon$ which is naturally small. The point is that ${\cal L}_0[\varphi]$ alone cannot generate radiative corrections which break the symmetry by assumption. In fact, symmetry breaking terms can only be generated by $O[\varphi]$ and must therefore be weighted by $\epsilon$.  It follows that radiative correction to $\epsilon$,  must be  $\lesssim {\cal O}(\epsilon)$, and so the scale of  $\epsilon$ can  be considered radiatively stable.

If we could find an enhanced  global symmetry in the limit of vanishing cosmological constant, then we could argue that any non-zero value of the cosmological constant was naturally small in the 'tHooft sense. Of course, finding such a symmetry that is compatible with Nature is the holy grail of particle physics

\subsection{Supersymmetry} Supersymmetry is a symmetry relating bosons and fermions. If supersymmetry is unbroken, a boson and its fermionic super partner share the same mass. This allows for a solution to the hierarchy problem, since the Higgs mass can be related to the mass of the Higgsino, which being fermionic is protected from large radiative corrections by chiral symmetry\cite{susy}.  Furthermore, from the point of view of the cosmological constant, we note that vacuum energy contributions from bosons enter with the opposite sign to fermions, and  supersymmetry enforces a cancellation i.e. 
\be
\underset{\text{boson loop}}{\unitlength=1mm
\begin{fmffile}{1loopvac1}
\parbox{15mm}{
 \begin{fmfgraph}(15,15)
    \fmfleft{i,i1,i2}
    \fmfright{o,o1,o2}
    \fmf{photon}{i2,v1}
    \fmf{plain,left=0.4}{v1,v2,v3,v4,v1}
    \fmf{phantom}{v2,o2}
     \fmf{phantom}{i,v4}
      \fmf{phantom}{v3,o}
  \end{fmfgraph}}
\end{fmffile}} +
\underset{\text{fermion loop}}{\unitlength=1mm
\begin{fmffile}{1loopvac1fermion}
\parbox{15mm}{
 \begin{fmfgraph}(15,15)
    \fmfleft{i,i1,i2}
    \fmfright{o,o1,o2}
    \fmf{photon}{i2,v1}
    \fmf{dashes,left=0.4}{v1,v2,v3,v4,v1}
    \fmf{phantom}{v2,o2}
     \fmf{phantom}{i,v4}
      \fmf{phantom}{v3,o}
  \end{fmfgraph}}
\end{fmffile}} =0
\ee
Suppose supersymmetry is spontaneously broken at a scale $M_{susy}$ then boson and fermion masses becomes non-degenerate by an amount $|M_{boson}^2-M_{fermion}^2| \sim g^2 M_{susy}^2$ where $g$ is the strength of their coupling to the supersymmetry breaking sector. Furthermore, the vacuum energy  cancellation fails by ${\cal O}(g^4 M_{susy}^4)$, so a technically natural value for the cosmological constant would be $\Lambda \sim g^4 M_{susy}^4$. Given the absence of any observed superpartner below the TeV scale \cite{RPP}, we infer that $gM_{susy} \gtrsim$ TeV, so in our Universe there just isn't enough supersymmetry to solve the cosmological constant problem.

\subsection{Scale invariance}  We already encountered scale invariance in the context of Weinberg's no go theorem. To see explicitly why it might help with the cosmological constant problem consider the Einstein equations sourced by the renormalised vacuum energy
\be
M_{pl}^2 G_{\mu\nu}=-\Lambda_{ren} g_{\mu\nu}
\ee
Now under a global scale transformation $g_{\mu\nu} \to \lambda^2 g_{\mu\nu}$, the Einstein tensor is invariant $G_{\mu\nu} \to G_{\mu\nu}$, in contrast to the source term $\Lambda_{ren} g_{\mu\nu} \to \Lambda_{ren} \lambda^2 g_{\mu\nu}$. This suggests that scale invariance should force the renormalised vacuum energy to vanish. Unfortunately there are a few problems with using scale invariance to tackle the cosmological constant problem.  Classical scale invariance is usually broken by quantum corrections. But even if scale invariance is preserved at the quantum level we know that it is broken in Nature, possibly spontaneously. By Weinberg's no go theorem we saw how  this means the effective potential must vanish at its minimum in order to preserve the translationally invariant vacuum. This vanishing of the effective potential at its minimum represents a fine-tuning that is spoilt by radiative corrections. In particular,
\be
V_{tree}(\varphi) |_{min}=0 \centernot \implies  V_{1loop} (\varphi)|_{min} =0
\ee
See \cite{Cliff} for a more extensive discussion of scale invariance in the context of the cosmological constant problem.

\subsection{Energy parity}
Consider an energy parity operator $P$ which reverses the expectation value of the Hamiltonian \cite{KS}. In other words, if $|\psi' \rangle=P | \psi \rangle$, then $\langle \psi' |H |\psi' \rangle=-\langle \psi |H |\psi \rangle$. If the vacuum is energy parity invariant then $|0' \rangle=|0\rangle$ and its energy must vanish since
\be
\langle 0|H |0  \rangle=-\langle 0|H |0  \rangle=0
\ee
Energy parity is the reason behind the vacuum energy cancellation in Linde's ingenious model of {\it Universe multiplication} \cite{linde2}.  The idea is simply to double up: our Universe is mirrored exactly by another Universe containing negative energies, with the two Universes interacting globally.  The action describing the model is given by
\be
S=\int d^4 x \int d^4 y \sqrt{-g(x)} \sqrt{-g(y)}\left[\frac{M_{pl}^2}{2} R(x)-{\cal L}_m[g^{\mu\nu}(x), \Psi(x)] -\frac{M_{pl}^2}{2}  R(y)+{\cal L}_m[g^{\mu\nu}(y), \Psi(y)]  \right]
\ee
where  the $x$ coordinates  correspond to our Universe, and the $y$ coordinates to its mirror.    The four dimensional metrics and corresponding Ricci scalars are given by $g_{\mu\nu}(x), R(x)$ and $g_{\mu\nu}(y),  R(y)$ respectively. Each Universe contains the same matter fields, described by $\Psi(x)$ and $\Psi(y)$. The action is manifestly invariant under energy parity provided the effective matter Lagrangians, ${\cal L}_m[g^{\mu\nu}(x), \Psi(x)] $  and ${\cal L}_m[g^{\mu\nu}(y), \Psi(y)] $ are the same in both Universes. The resulting field equations are given by
\be
M_{pl}^2 G_{\mu\nu}(x)=T_{\mu\nu}(x)-\Lambda_x g_{\mu\nu}(x), \qquad M_{pl}^2 G_{\mu\nu}(y)=T_{\mu\nu}(y)-\Lambda_y g_{\mu\nu}(y)
\ee
with 
\begin{eqnarray}
\Lambda_x &=& \frac{\int d^4 y  \sqrt{-g(y)} \left[ \frac{M_{pl}^2}{2}  R(y)-{\cal L}_m[g^{\mu\nu}(y), \Psi(y)] \right]}{\int d^4 y  \sqrt{-g(y)}} \\
\Lambda_y &=& \frac{\int d^4 x  \sqrt{-g(x)} \left[ \frac{M_{pl}^2}{2}  R(x)-{\cal L}_m[g^{\mu\nu}(x), \Psi(x)] \right]}{\int d^4 x  \sqrt{-g(x)}}
\end{eqnarray}
Thus the cosmological counterterm in our Universe, $\Lambda_x$, is given by the historic average of fields in the mirror Universe (and vice versa).  Since the effective matter Lagrangians in each Universe are assumed to be identical\footnote{Not only do we assume the same matter content, but also the same cut-off and the same order in loops.}, they contribute the same vacuum energy
\be
\langle 0 |  {\cal L}_m[g^{\mu\nu}(x), \Psi(x)]  | 0 \rangle = \langle 0 |  {\cal L}_m[g^{\mu\nu}(y), \Psi(y)]  | 0 \rangle=V_{vac}
\ee
The historic average of a constant is just the same constant, and so it follows that the vacuum energy in the mirror contributes $-V_{vac}$ to the counterterm $\Lambda_x$. This precisely balances the contribution to the energy momentum tensor of the vacuum energy in our Universe, $T_{\mu\nu}^{vac}(x)=-V_{vac} g_{\mu\nu}(x)$.  The vacuum energy  drops out of the dynamics of our Universe completely, and by the same mechanism it also drops out of the dynamics of the mirror.

Any attempt at quantising this theory will presumably  run into serious difficulties on account of the fact ordinary fields    in our Universe must be reflected by ghosts in the mirror, with the two systems interacting globally.  There are other concerns too: the cancellation of vacuum energy relies heavily on the assumption that the effective matter Lagrangians in each Universe are  identical. Is this justified? Furthermore, the evolution of our Universe is highly sensitive to any change in the evolution of the mirror. 

Energy parity symmetry was also explored relatively recently in a (somewhat speculative) model due to Kaplan and Sundrum \cite{KS}. In the absence of gravity the Lagrangian for ordinary matter is balanced by a ghostly counterpart with opposite sign. At this level the ghost is safe because there is no direct interaction between the two sectors. Such a theory is energy parity invariant, with positive energies in one sector mirrored by negative energies in the other resulting in cancellation of vacuum energy.  When gravity is switched on energy parity is broken and quantum gravity loops can contribute a vacuum energy of order $\mu^4$ where $\mu \sim$ meV is the cut off of the effective gravity theory. The graviton also helps to mediate an interaction between the two sectors leading to a catastrophic instability in a Lorentz invariant theory. The claim is that this should be tamed by Lorentz violating effects above the cut-off $\mu$.

Finally, note that a ``complex transformation", $x^\mu \to i x^\mu$, acts like a stress energy parity operator, since $T_{\mu\nu} \to -T_{\mu\nu}$ \cite{tHN} (see also \cite{ds-ds}). This naturally relates de Sitter space to anti de Sitter space, and only remains unbroken for vacua with vanishing cosmological constant. 'tHooft and Nobbenhuis \cite{tHN} suggest that such a complex   symmetry may only be broken by boundary conditions on physically excited states (much in the same way that translational invariance is often only broken by the boundary).  The symmetry remains unbroken by the vacuum, thereby forcing a vanishing cosmological constant. However, it is difficult to incorporate non-vanishing masses into this scenario (since $p^2  \to -p^2$), as well as non-abelian gauge theories \cite{tHN}.

\section{Colemania}  \label{sec:cole}

Coleman's theory of the cosmological constant \cite{cole} may not have stood the test of time, but it was remarkably clever, and worthy of inclusion in any review of the cosmological constant problem. In his original paper, Coleman speaks of {\it precognition}, with  the Universe knowing all along that  it had to grow old and big.  This resonates with much of what I will say in section \ref{sec:long}, so it is instructive to understand where the theory  went right, and where it went wrong. To this end, let us return to the eighties and consider the path integral for Euclidean quantum gravity.  What effect do Euclidean wormhole solutions \cite{strom} have on the path integral? If we consider wormholes that are small but subPlanckian,  and integrate them out, we find that they modify  coupling constants, $\lambda$, and provide a probability function, $P(\lambda)$ \cite{cole}. To see how this transpires we shall follow the review of Coleman's mechanism presented in \cite{Kleb}.

\begin{figure}[h]
\begin{center}
\epsfig{file=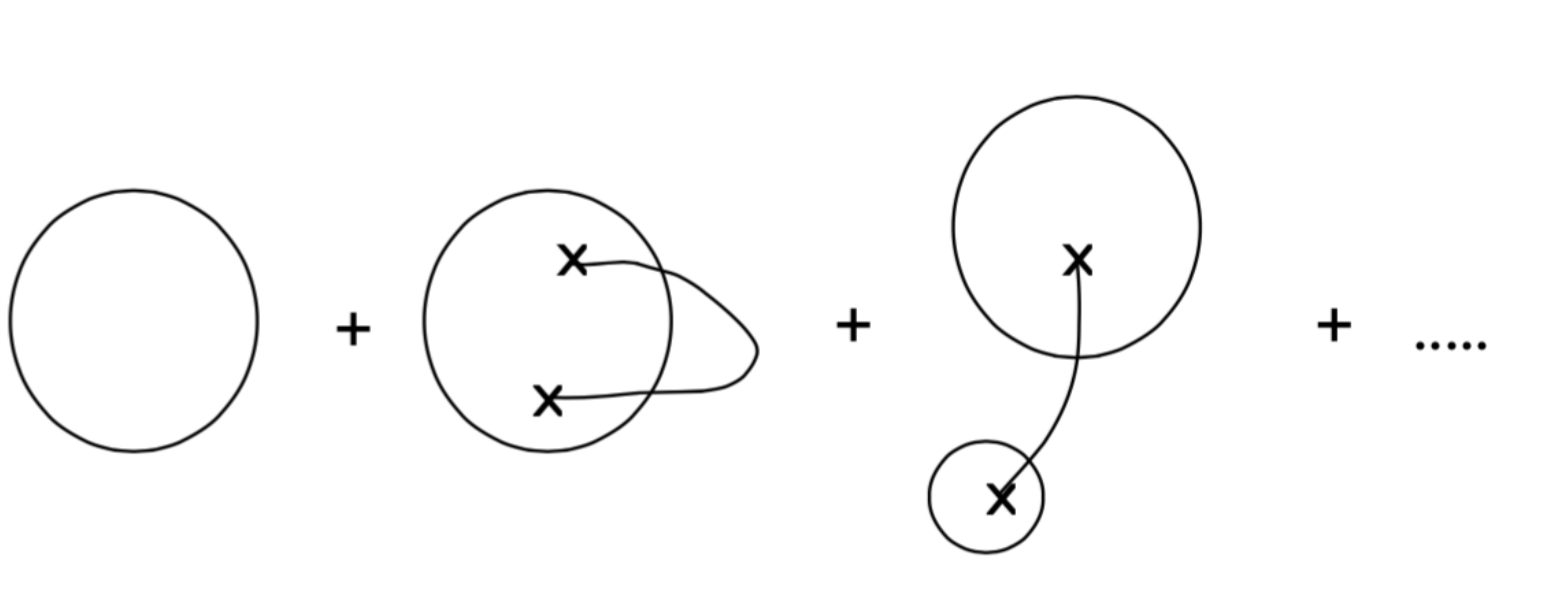, width= 120 mm, height= 50 mm}
\caption{Wormhole corrections to the Euclidean path integral, including wormholes that connect points within the same universe and different universes} \label{fig:wh}
\end{center}
\end{figure}

Our goal is to integrate out wormholes and field fluctutations of size $L$ where $M_{pl}^{-1}<L< \mu^{-1}$, for some scale $\mu$. We start with the wormholes. In their absence, the expectation value of an operator ${\cal Q}$  is expressed as
\be \label{wheq}
\langle {\cal Q} \rangle_\lambda=\frac{\int D g {\cal Q} e^{-I[g, \lambda)}}{\int D g e^{-I[g, \lambda)}}
\ee
where $I[g, \lambda)$ is the Euclidean action for fields $g$ (including metric and matter) with coupling constants, $\lambda$. We are interested in wormhole corrections like those shown in Figure \ref{fig:wh}.
The wormholes are assumed to be dilute, so that their average separation is greater than their size, and there are two types: wormholes connecting distant points in a single universe, and wormholes connecting different universes.  We shall build the wormhole corrected path integral  bit by bit, correcting first for the single universe wormholes, and then their multi-universe counterparts.

To include a wormhole connecting points $x$ and $x'$ in the same Universe we insert an operator $ C_{ij} \O_i(x) \O_j(x')$, where $C_{ij} \sim e^{-S_W}$ and $S_W$ is the wormhole action.  Here the $\O_i(x)$ are a basis of local operators, and the $C_{ij}$ are treated as constant above the wormhole scale. Repeated indices are understood to be summed over. If we sum over the ends of the wormhole  the numerator in (\ref{wheq}) is replaced by
\be \label{wheq1}
\int D g {\cal Q} e^{-I[g, \lambda)} \frac12 \int dx \int dx' C_{ij} \O_i(x) \O_j(x')
\ee
Now we include any number of wormholes connecting a single universe, so that the sum  exponentiates and (\ref{wheq1}) becomes
\be \label{wheq2}
\int D g {\cal Q} e^{-I[g, \lambda)+\frac12  \int dx \int dx'  C_{ij} \O_i(x) \O_j(x')}
\ee
This can be rewritten as 
\be \label{wheq3}
 \int d \alpha  \int D g    {\cal Q}  e^{-\left[I[g, \lambda)+\frac12  D_{ij} \alpha_i \alpha_j+\alpha_i \int d^4 x \O_i(x) \right]}
\ee
where $D_{ij}=C_{ij}^{-1}$ and  we have made use of the following identity
\be
e^{\frac12 C_{ij} V_i V_j}=\int d \alpha e^{-\frac12 D_{ij} \alpha_i \alpha_j-\alpha_i V_i}
\ee
with $V_i=\int d^4x \O_i(x)$. Without loss of generality, we can define the $\lambda_i$ to be the coefficient of the operator $\O_i$ in the Lagrangian, so that we can rewrite (\ref{wheq3}) as
\be \label{wheq4}
 \int d \alpha  \int D g    {\cal Q}  e^{-\left[I[g, \lambda+\alpha)+\frac12  D_{ij} \alpha_i \alpha_j \right]}
\ee
Thus one effect of the wormholes is to shift the fundamental coupling constants, as we originally claimed.

Now we include the multi-universe contribution, when wormholes connect us to  additional closed universes. This brings in new factors of 
\be
\int Dg'  e^{-I[g', \lambda+\alpha)}=Z(\lambda+\alpha)
\ee
so that upon summing over many additional closed universes, the sum exponentiates and we find that 
\be \label{wheq5}
\langle {\cal Q} \rangle=\frac{1}{N}  \int d\alpha \int D g {\cal Q} e^{-\left[I[g, \lambda+\alpha)+\frac12  D_{ij} \alpha_i \alpha_j \right]+Z(\lambda+\alpha)}
\ee
where  $N$ is the normalisation constant.  Let us now explicitly integrate out heavy field fluctuations, above the wormhole scale, $\mu$. To this end we split $g$ up in terms of  light modes  $g_l$ and heavy modes  $g_h$,  then integrate out the latter by writing  (\ref{wheq5}) as 
\be \label{wheq6}
\langle {\cal Q} \rangle=\frac{1}{N} \int d\alpha  \int D g_l {\cal Q} e^{-\left[I_\text{eff}[g_l, \lambda(\mu)+\alpha)+\frac12  D_{ij} \alpha_i \alpha_j \right]+Z_\text{eff}(\lambda(\mu)+\alpha)}
\ee
where $I_\text{eff}[g_l, \lambda(\mu)+\alpha)$ is an effective action for the lighter modes in a wormhole-free universe, 
\be
 e^{- I_\text{eff}[g_l, \lambda(\mu)+\alpha)}=\int D g_h e^{-I[g_l, g_h, \lambda(\mu)+\alpha)}
\ee 
and 
\be
Z_\text{eff}(\lambda(\mu)+\alpha)=\int Dg_l e^{- I_\text{eff}[g_l, \lambda(\mu)+\alpha)}=Z(\lambda+\alpha)
\ee
Obviously we expect the  couplings $\lambda(\mu)$  to run in the usual way as we integrate out the heavy field modes. 

Now we could use  (\ref{wheq6}) to define a wormhole corrected effective action,
\be
{\cal I}_\text{eff}[g_l, \alpha)=I_\text{eff}[g_l, \lambda(\mu)+\alpha)+\frac12  D_{ij} \alpha_i \alpha_j -Z_\text{eff}(\lambda(\mu)+\alpha)
\ee
satisfying $\langle {\cal Q} \rangle=\frac1N  \int d\alpha  \int D g_l   {\cal Q}   e^{-{\cal I}_\text{eff}[g_l, \alpha)}$.  Casting our eyes towards section \ref{sec:long}, we note that this looks a lot like a sequestering action (\ref{seqact}), with functions of global variables, $\alpha$, sitting outside of the integral. Indeed, if we were to proceed as in the sequester, we would  use the method of steepest descent to extract the following classical equations of motion,
\be \label{seqwh}
\frac{\delta  {\cal I}_\text{eff}}{\delta g_l}=0, \qquad \frac{\del  {\cal I}_\text{eff}}{\del \alpha_i}=0
\ee
with the latter corresponding to global constraints. Coleman takes a different (and not equivalent) route. He essentially writes equation (\ref{wheq6}) as
\be \label{wheq6}
\langle {\cal Q} \rangle=\int d\alpha \langle {\cal Q} \rangle_{\lambda(\mu)+\alpha} P(\alpha)
\ee
extracting a probability density
\be \label{Pa}
P(\alpha)=\frac1N  e^{-\-\frac12  D_{ij} \alpha_i \alpha_j +Z_\text{eff}(\lambda(\mu)+\alpha)}
\ee
for the coupling constants. In other words, the expectation value of ${\cal Q}$ is the weighted average of the expectation value of ${\cal Q}$ in wormhole-free universes (in the effective theory). The probability density depends on the form of the wormhole propagator $D_{ij}$ and the path integral, $Z_\text{eff}(\lambda(\mu)+\alpha)$, for a wormhole-free Universe with shifted couplings.  To approximate the form of $P(\alpha)$, Coleman notes that, for gravity, we expect
\be
 I_\text{eff}=-\int d^4 x \sqrt{g} \left[-\Lambda(\mu)+ \frac{R}{16 \pi G(\mu)}  +\xi(\mu)(\text{Riemann})^2 +\ldots \right]
\ee
where $\Lambda(\mu), G(\mu)^{-1}, \xi(\mu), \ldots$ are identified with the shifted coupling constants $\lambda(\mu)+\alpha$. He  then assumes that Einstein gravity dominates the path integral, so that we have
\be
R_{\mu\nu}=8 \pi G \Lambda g_{\mu\nu}
\ee
with the the large $4$-sphere (Euclidean de Sitter) of radius $r=(3/8\pi G\Lambda)^{1/4}$ being the dominant solution. Plugging this into the effective action yields
\begin{multline}
I_\text{eff} |_\text{$4$-sphere}= -\int d^4 x \sqrt{g} \left[ \Lambda + {\cal O}(1) \xi (8 \pi G \Lambda)^2 +\ldots\right]\\ = -\frac83 \pi^2 r^4 \left[ \Lambda + {\cal O}(1) \xi (8 \pi G \Lambda)^2 +\ldots\right]  = -\left[ \frac{3}{8 G^2 \Lambda}+\xi{\cal O}(1) +\ldots \right] 
\end{multline}
and the following approximation to the wormhole-free path integral
\be
Z_\text{eff}(\lambda+\alpha) \approx e^{-I_\text{eff} |_\text{$4$-sphere} } =e^{\frac{3}{8 G^2 \Lambda}+\xi{\cal O}(1) +\ldots}
\ee
Given the form of the probability density $P(\alpha)$,  this suggests the probability  is peaked near $G^2\Lambda=0^+$, and so $\Lambda$ should be tiny in Planck units. As Coleman says, this is why we have nothing rather than something.

Unfortunately, Coleman's proposal suffers from a catastrophe \cite{FS}\footnote{Kaplunovksy is also credited with having identified this problem in private communications.}. In our presentation, we deliberately included the $(\text{Riemann})^2$ piece in the effective action so as to demonstrate this explicitly.  Indeed, as Weinberg \cite{Wein} first noted, the probability density is also peaked at $|\xi| \to \infty$, which would be in complete violation of perturbative unitarity. He therefore proposed a unitarity bound of $\xi < \frac{1}{G\mu^2}$ for an effective theory cut-off at the scale $\mu$. The wormhole catastrophe \cite{FS} now arises as we change the cut-off $\mu \to \mu'< \mu$. Of course, it is not enough just to integrate out the additional fluctuations in the fields $g$. We also need to integrate out those wormholes of size, $L'$ where $\mu^{-1} <L'<\mu'^{-1}$. Their effect is to add in new auxiliary variables, $\alpha'$, analogous to $\alpha$, so that (\ref{wheq6}) becomes
\be \label{wheq6}
\langle {\cal Q} \rangle=\frac{1}{N} \int d\alpha  \int D g'_l {\cal Q} e^{-\left[I'_\text{eff}[g'_l, (\lambda+\alpha)(\mu, \mu')+\alpha')+\frac12  D_{ij} \alpha_i \alpha_j +\frac12  D'_{ij} \alpha'_i \alpha'_j \right]+Z'_\text{eff}((\lambda+\alpha)(\mu, \mu')+\alpha')}
\ee
where $g'_l$ denote the ultralight field modes below the scale $\mu'$, and $I'_\text{eff}$ the new effective action on a wormhole-free universe. Note that in integrating out the  field modes with intermediate energies, the small-wormhole shifted couplings $\lambda(\mu)+\alpha$ run  to some $(\lambda+\alpha)(\mu, \mu')$.  Integrating out the intermediate wormholes introduces an additional shift $\alpha'$ in the renormalised coupling.

Working through a similar argument we now impose a weaker unitarity bound on the $(\text{Riemann})^2$ couplings, $\xi < \frac{1}{G\mu'^2}$.  This suggests that the wormhole contributions to $\xi$ increase  at large distances.  This is the wormhole catastrophe: death by giant wormholes.  Fischler and Susskind further show that  the density of wormholes at the scale $\mu$ scales as $1/\mu^4$.  Thus the density of large wormholes is high, which makes some intuitive sense: large wormholes increase the volume of space on which to attach more wormholes.

\section{Short distance modifications of gravity}  \label{sec:short}
General Relativity is not perturbatively renormalisable, and for that reason we expect new gravitational physics to kick in at short distances. Furthermore, the short distance limit of tabletop gravity tests $ \sim 1/\text{meV}$ happens to coincide with the dark energy scale $\rho_{DE} \sim (\text{meV})^4$. Could  new gravitational physics above meV  possibly help with the cosmological constant problem?  We suspect not. Gravitons probing dark energy are ultrasoft, with wavelengths of order the current Hubble distance, $\sim 1/H_0$. How could they ever be affected by new physics at a meV? That said, interesting short distance modifications of gravity have been proposed and we will review them presently.

\subsection{Supersymmetric large extra dimensions (SLED)} Over the last ten years or so, the SLED proposal \cite{SLED} has possibly been the most popular candidate for solving the cosmological constant problem. The most recent manifestation of the model consists of six dimensional Nishino-Sezgin supergravity \cite{NS}, with our Universe corresponding to one of two non-supersymmetric $3$-branes. The supersymmetric ``bulk" is compactified in the shape of a rugby ball and has the following bosonic field content:  a metric, $g_{ab}$, a dilaton, $\phi$, and a $U(1)_R$ gauge potential, $A_a$.  To get a feel for how the model works we shall now go through each of the letters in the SLED acronym. The reader interested in a more detailed description of the model should  begin with Burgess' review \cite{Cliff}, and go from there. 

\paragraph{Extra dimensions (ED)} The SLED proposal exploits the extra dimensional loophole: four dimensional vacuum energy  does not have to generate four dimensional curvature. It can curve the extra dimensions.  Indeed, imagine we have $n$ extra dimensions. Four dimensional vacuum energy now corresponds to the tension of a $3$-brane.  Focussing purely on the metric sector for simplicity, the action in the vicinity of a single brane is given by
\be S=\frac{1}{16 \pi G_{n+4}} \int d^{n+4} x \sqrt{-g_{n+4}}  R-T_{b} \int d^4 x \sqrt{-g_4} 
\ee
The field equations should balance the brane tension against a bulk singularity such that
\be
G_{ab}^{singular}=-8 \pi G_{n+4} T_{b} \frac{\delta^{(n)}(x-x_{brane})}{\sqrt{g_n} }g_{4\mu\nu}\delta^\mu_a \delta^\nu_b
\ee
Plugging this singular piece back into the action, we can construct an effective brane tension given by
\be
T_b^\text{eff}=T_b -\frac{1}{16 \pi G_{n+4}} \int d^{n} x \sqrt{-g_{n}}  R^{singular}=T_b \left( \frac{n-2}{n+2}\right)
\ee
For $n=2$ extra dimensions, we see that the effective tension vanishes. This should not come as any surprise. In four dimensions, the tension of a cosmic string merely creates a deficit angle in the bulk, rather than curving the string itself. In the SLED model, the same phenomena allows us to have flat brane solutions for any value of the brane tension.  Two extra dimensions also means the extra dimensions can be as large as a millimeter\footnote{Note the effective four dimensional Newton's constant, $G_4 \sim G_{n+4}/l^n$, where $l$ is the size of the extra dimensions. Given that $G_4 \sim (1/10^{18} \text{GeV})^2$ and requiring $G_{n+4} \lesssim (1/\text{TeV})^{n+2}$ to avoid quantum gravity having already shown up in accelerators, we extract the  bound $l \lesssim 10^{\frac{30}{n}-15}$ mm, and so $l \sim mm \implies n \leq 2$.}, which will be useful later. 

\paragraph{Supersymmetric (S)} The claim is that a flat brane solution is guaranteed by the SLED model provided (1) the brane stress-energy vanishes in the off-brane directions; and (2) brane-dilaton couplings are absent. Generically these two criteria are broken by loops, but if the scale of supersymmetry breaking in the bulk is low enough, the effects are said to be small.

\paragraph{Large (L)} Large extra dimensions are important because explicit calculations show that the effective four dimensional vacuum energy goes like the Kaluza-Klein scale, i.e. $V_{vac}^\text{eff} \sim 1/l^4$. This means the extra dimensions should be no smaller than a millimetre $\sim 1/$meV. Table top tests of gravity require them to be no larger than a millimetre, and as we have seen, $n=2$ allows precisely millimetre sized extra dimensions.  
\newline

Although it is not deemed important enough to appear in the acronym, bulk scale invariance also plays a crucial role, helping to suppress the bulk cosmological constant at tree level. Note also that the model is not in conflict with Weinberg's no go theorem. Below the scale of compactification, $1/l$, we would expect Weinberg's theorem to prohibit any cancellation of vacuum energy without fine-tuning. In SLED there is no cancellation below $1/l$, but that doesn't matter because $l$ is so large. Despite the claim of success, however, there remain concerns as to whether gauge invariant bulk solutions with finite extra dimensional volume really do evade  fine tuning \cite{private}.

\subsection{Fat graviton}
\begin{figure}[h]
\begin{center}
\epsfig{file=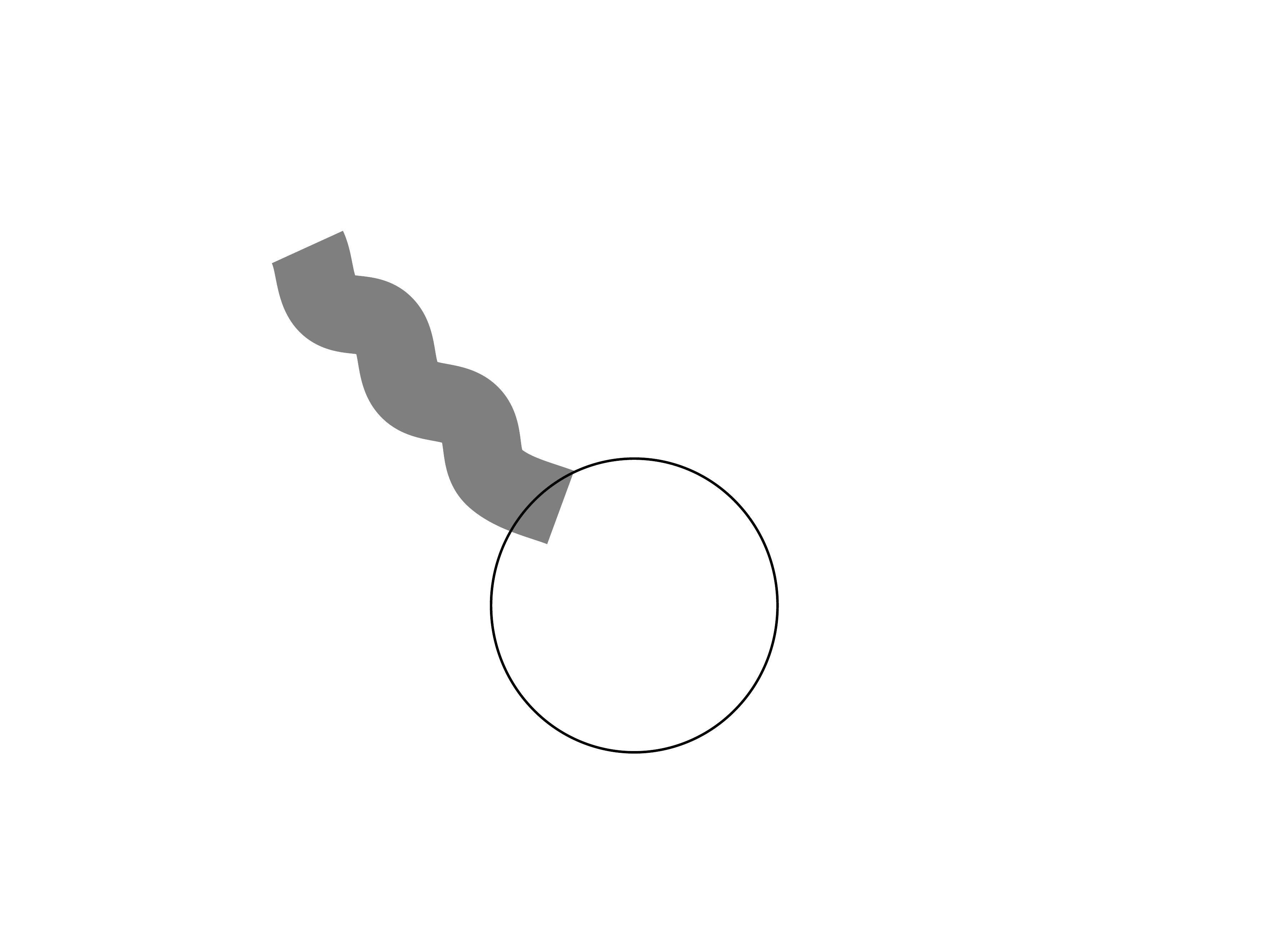, width= 70 mm, height= 50 mm}
\epsfig{file=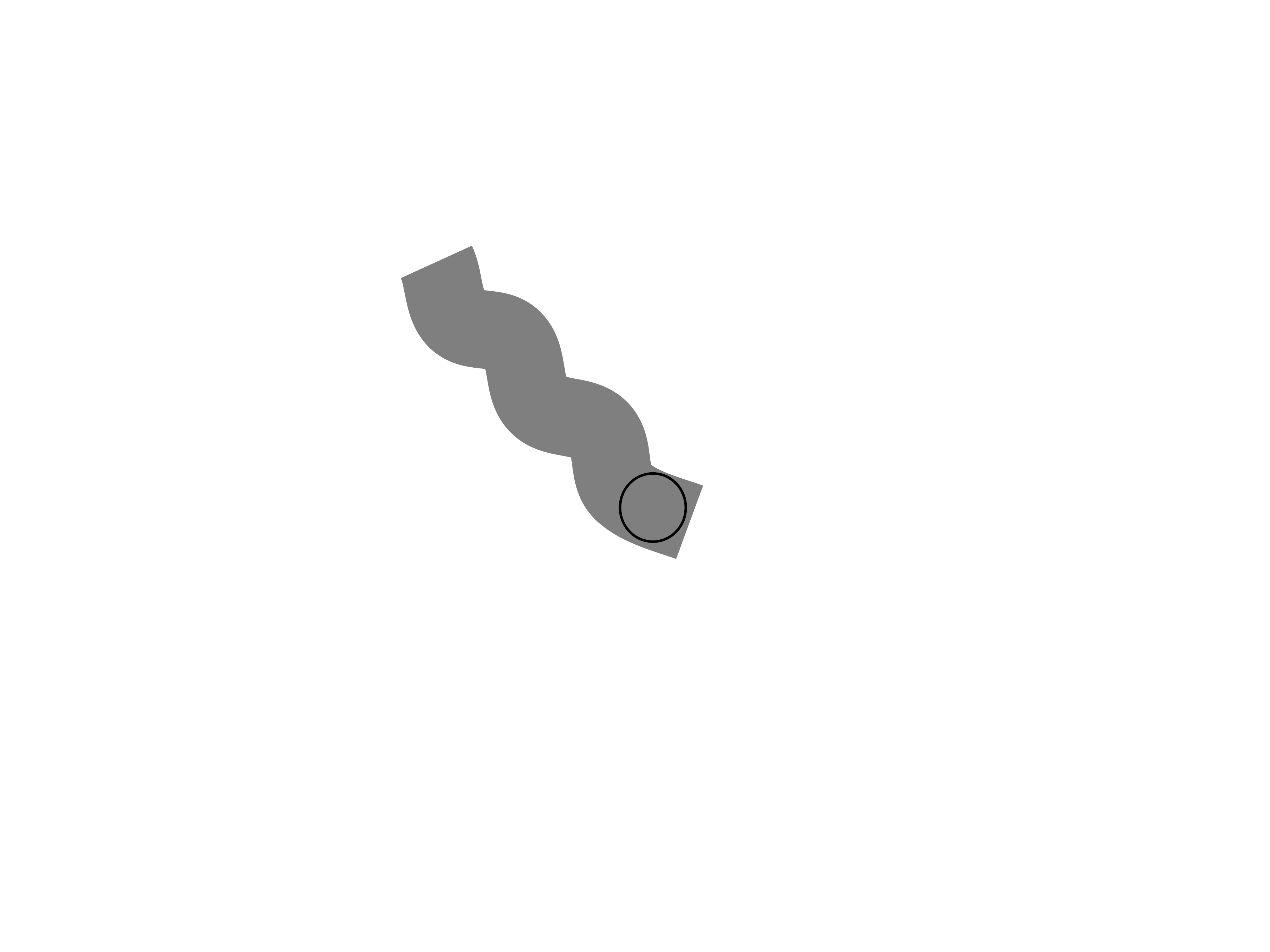, width= 70 mm, height= 50 mm}
\caption{Position space Feynman diagrams with a Standard Model loop and external fat graviton. The left hand diagram shows a soft (i.e. large) Standard Model loop. This behaves just like the familiar tadpole diagram in GR  and therefore makes a contribution to the vacuum energy.  The right hand diagram shows a hard (i.e. small) Standard Model loop which is completely swamped by the fat graviton. This is assumed to be suppressed,  making no contribution to the vacuum energy.} \label{fig:fat}
\end{center}
\end{figure}

The basic idea behind the fat graviton is to give the graviton a width  $\sim 1/$meV \cite{fat}. Soft  Standard Model couplings to the graviton are point like, as in General Relativity. In contrast, hard Standard Model couplings  to the graviton are {\it not} point like but global and suppressed. Because of this  we are able to throw away the contributions to vacuum energy from heavy Standard Model fields. This  is demonstrated pictorially in figure \ref{fig:fat}.
To develop the idea, Sundrum constructs an effective effective field theory of soft gravitons, and includes the effects of heavy Standard Model fields using methods familiar from Heavy Quark Effective Field Theory. It turns out that the largest contribution to the cosmological constant comes from vibrational modes of the fat graviton, scaling like $(1/\text{graviton width})^4$.  

Although hard Standard Model loops are suppressed in diagrams such as the one on the right hand side of figure \ref{fig:fat}, if external Standard Model particles are included, there is no suppression. We know this has to be true from our discussion of the Lamb shift. The rules of suppression, or non-suppression, seem to have been imposed by fiat at the level of the effective field theory. Such considerations will presumably  make it very hard to realise the desired UV properties of the fat graviton theory from a local theory.

\section{Long distance modifications of gravity}  \label{sec:long}
There is a clear sense in which the cosmological constant problem can be thought of as an infra-red issue.  The observational constraints come from ultrasoft gravitons probing dark energy at Hubble distances. We can also think of  the  cosmological constant as the gravitational source of longest wavelength. For these reasons we have seen a surge in attempts to modify General Relativity at very large distances (for recent reviews see \cite{review1,review2}).  Indeed, the basic idea behind {\it degravitation} \cite{degrav,degrav2} is to have gravity ``switch off' beyond some large scale $L$, so that ultra long wavelength sources like the cosmological constant no longer gravitate.

Modifications of gravity at some large but finite scale are prevalent in the literature. A notable example is massive gravity \cite{FP} and I refer the reader to  \cite{kurtreview,dRreview} for extensive reviews. Another example is  the Fab Four \cite{fab4} which will be discussed in more detail shortly.  I should emphasise, however, that modifying gravity at a large but finite scale is notoriously tricky, mainly due to the phenomenological success of General Relativity at much shorter distances. The point is that a significant deviation from GR at Hubble scales must have gone away to one part in $10^5$ by the time we reach the solar system\cite{Will}. This necessitates some sort of {\it screening mechanism} to suppress the effect of the modification in the gravitational field of the Sun and the Earth.  Proposals include the chameleon \cite{cham}, the symmetron \cite{symmetron} and the Vainshtein mechanism \cite{vain}.  The latter is probably the most efficient, but is not without its problems, in particular issues with UV completions \cite{vainUV}.  In the context of the cosmological constant problem, there is also an  intimate connection between long distance modifications of gravity and  violations of causality \cite{degrav}.

In these lectures, I would like to advocate the ultimate long distance modification of gravity - a {\it global} modification of gravity. This has several advantages, not least the fact that locally such a theory is equivalent to General Relativity so there is no need to jump through hoops to pass solar system tests.  A global modification also cuts straight to the core of the cosmological constant, which we can think of as the global component of the energy momentum tensor.  The sequestering scenario \cite{KP1,KP2,KP3} represents an example of a global modification of gravity in which the radiative instability of the Standard Model vacuum energy may be eliminated. Most of this section will be devoted to this model.

\subsection{The Fab Four}
The Fab Four model \cite{fab4} grew from an attempt to generalise Weinberg's no go theorem, relaxing the assumption of translation invariance at the level of the self-adjusting fields. In other words, allow the self-adjusting fields to be non-constant, demanding only a constant (i.e. Minkowski) metric. 

The simplest example of a self-adjusting field is, of course, a single scalar, $\phi$.  Because the most general Lorentz invariant  theory of a tensor coupled to a single scalar degree of freedom was written down long ago by Horndeski\footnote{Horndeski's theory has received a renaissance since being rediscovered in the first Fab Four paper. Horndeski himself abandoned mathematics long ago and has since become an artist. The story goes that he went to Holland for a conference and took time out to visit a van Gogh exhibition. He was so inspired that he quit mathematics and took to painting himself. See \cite{hornwork} for examples of his artwork.}  \cite{horn} (see also \cite{cedric}), one can now ask the following question: suppose I take Horndeski's theory, is there any corner of the theory that can successfully self-adjust if I allow the scalar to vary over time? More precisely, let me assume that matter is minimally coupled to the metric only.  Is there any subtheory of Horndeski that admits a Minkowski solution for {\it any} value of the vacuum energy, even allowing for phase transitions?  The answer turns out to be ``yes" and the corresponding subtheory is described by the Fab Four action,
\begin{multline}
S_{\text{Fab4}}[g_{\mu\nu}, \phi]=\int d^4 x \sqrt{-g}\left[ V_{john}(\phi) G_{\mu\nu}\nabla^\mu \phi \nabla^\nu \phi +V_{paul}(\phi) P_{\mu\nu\alpha\beta} \nabla^\mu \phi \nabla^\alpha\phi \nabla^{\nu} \nabla^\beta \phi \right. \\
\left.  +V_{george}(\phi) R+V_{ringo}(\phi) R_{GB}\right]
\end{multline}
where $P^{\mu\nu\alpha\beta}$ is the double dual of the Riemann tensor, and $R_{GB}$ is the Gauss-Bonnet combination \cite{mtw}. In order to evade Weinberg's theorem, the scalar itself must evolve, allowing the scalar equation equation of motion to force the vacuum curvature to zero, even in the presence of a large vacuum energy. Indeed, the vacuum energy sets the scale over which the scalar changes. At shorter scales Weinberg's theorem does apply, but in this limit the vacuum energy is small compared to the energy scales being probed, and we may think of this as the scale invariant limit rather than fine-tuning.

The presence of a light scalar suggests the need for some sort of screening mechanism in order that the Fab Four can pass solar system gravity tests\footnote{Recently it was shown that a generalisation of the Fab Four with low scale de Sitter self-tuning admitted de Sitter Schwarzschild as a solution even in the presence of a large vacuum energy \cite{boss}.}.  As we alluded to earlier, screening mechanisms may bring in other worries such as a low effective theory cut-off, and problems with superluminality. Such considerations were  explored in \cite{nkf4} in the absence of a cosmological constant, eliminating some, but not all, possibilities.  Furthermore, it is by no means  guaranteed that the structure of the Fab Four is stable against radiative corrections. 
\subsection{Long distance  modification of gravity and causality}
If we are to solve the cosmological constant problem through a long distance modification of gravity, we are likely to be challenged by causality \cite{degrav}. To see this, imagine a local theory which respects causality  and for which gravity shuts down at distances $L \gtrsim 1/H_0$. Just after the big bang, the energy momentum tensor contains both short and long wavelength contributions
\be
T_{\mu\nu}=T_{\mu\nu}^{short}+T_{\mu\nu}^{long}
\ee
with the distinction between long and short corresponding to the scale $L$. The long wavelength sources get eaten by the modification of gravity, by assumption. Now, in a causal theory, we have wait until a time $t \gtrsim L$ before  we can establish what sources live in $T_{\mu\nu}^{long}$, and what live in $T_{\mu\nu}^{short}$? This means the theory cannot decide which sources to degravitate early on, and so there are now three  possibilities to consider :
\begin{enumerate}
\item the large vacuum energy does not get cancelled until late on. 
 \item short wavelength sources are also cancelled along with those of long wavelength. 
\item causality should be abandoned.
\end{enumerate}
The first possibility is immediately ruled out by observation since the presence of a large cosmological constant right up until the current epoc would completely ruin the success of nucleosynthesis. The second possibility means that the theory isn't choosy: vacuum energy is cancelled by the modification, along with a bunch of short distance modes. This is extremely dangerous, and suggests that the phenomenology of short distance gravity will  be problematic\footnote{The Fab Four probably falls into this category.}. The third and final possibility seems like sacrilege. Or is it?

Certainly a local violation of causality should be avoided, but what if causality is only violated globally through a future boundary condition? Then there are no closed timelike curves and none of the pathologies one would usually associate with acausality.  Actually, the moment we define a black hole event horizon in General Relativity we violate causality in this way, making reference to future null infinity. 

We can sum up the gist of this argument by asking   how we should actually measure the zero point energy we are trying to cancel.  We cannot measure it just by looking within, say, our Hubble volume  since we cannot be sure that what we are measuring within that volume is not some locally flat section of a non-constant potential. Even if we have access to the whole of space, we cannot the measure the zero point energy over the course of a Hubble time since we cannot be sure that the potential won't change at some later time.  The thing about the zero point energy is that it is constant, over all of space and all of time, and to truly measure it we need to scan the whole of spacetime. We need to go global. These considerations point towards a global modification of gravity, which brings us to the sequester \cite{KP1,KP2, KP3}.

\subsection{The Sequester}
Consider the action for classical General Relativity  coupled to a quantum matter sector (\ref{GRact}),  assumed to contained the Standard Model, and imagine promoting the cosmological ``counterterm", $\Lambda$, to that of a global dynamical variable. That is not to say $\Lambda$ is a field that varies in space and time. It does not. It is a spacetime constant, but we can vary over it in the action. What we would really like to do is to get the variable $\Lambda$ to talk to the Standard Model vacuum energy\footnote{Since we are assuming classical gravity, we will not consider graviton loop contributions to the vacuum energy.} in just the right way as to cancel it.  To this end, we introduce a second global dynamical variable, $\lambda$, that  knows about the scales, and therefore vacuum energy, in a ``protected"  matter sector taken to include the Standard Model.  We then get the two dynamical variables to talk to one another by including a global interaction whose form is set by dimensional analysis. What follows is the {\it sequestering} action  \cite{KP1,KP2}
\be \label{seqact}
S=\int d^4 x \sqrt{-g} \left[ \frac{M_{pl}^2}{2}R -\lambda^4 {\cal L}_m(\lambda^{-2} g^{\mu\nu}, \Psi) - \Lambda \right]+\sigma \left(\frac{\Lambda}{\lambda^4 \mu^4}\right)
\ee
where $\mu$ is some mass scale.  Note that the global interaction $\sigma \left(\frac{\Lambda}{\lambda^4 \mu^4}\right)$ is not integrated over. The  precise form of the sequestering function $\sigma$ should be determined by phenomenology, although we do require it to be odd and differentiable. The global parameter $\lambda$ fixes the hierarchy between matter scales and the Planck mass. To see this note that the matter Lagrangian can be written as
\be
\sqrt{-g}  \lambda^4 {\cal L}_m(\lambda^{-2} g^{\mu\nu}, \Psi) =\sqrt{-\tilde g} {\cal L}_m(\tilde  g^{\mu\nu}, \Psi)
\ee
where $\tilde g_{\mu\nu}=\lambda^2 g_{\mu\nu}$.  This means that $\lambda$ is nothing more than a constant rescaling between the tilded ``Jordan" frame and the untilded ``Einstein" frame.  The Planck mass is fixed in the latter, while matter scales go like $m = \lambda \tilde m$, with $\tilde m$ the bare mass scale appearing in the matter Lagrangian.  Note that there is some conceptual overlap between the sequester and an old model of Tseytlin \cite{Tseytlin} (see also, \cite{others}), although it is only the sequester that achieves vacuum energy cancellation beyond tree-level.

Variation of the action (\ref{seqact}) with respect to the global variables $\Lambda$ and $\lambda$, and the local variable, $g_{\mu\nu}(x)$ yields the following equations of motion
\begin{eqnarray}
\frac{1}{\lambda^4 \mu^4} \sigma' \left(\frac{\Lambda}{\lambda^4 \mu^4}\right) &=& \int d^4 x\sqrt{-g}  \label{lambda}\\
\frac{4\Lambda}{\lambda^4 \mu^4} \sigma' \left(\frac{\Lambda}{\lambda^4 \mu^4}\right) &=& \int d^4 x\sqrt{-g} \lambda^4 \tilde T^\alpha_\alpha \label{Lambda} \\
M_{pl}^2 G^\mu_\nu &=&-\Lambda \delta^\mu_\nu +\lambda^4 \tilde T^\mu_\nu \label{g}
\end{eqnarray}
where $\tilde T_{\mu\nu}=\frac{2}{\sqrt{-\tilde g}} \int d^4 x \sqrt{-\tilde g} {\cal L}_m(\tilde  g^{\mu\nu}, \Psi)$.  The last equation is just the equation of motion of General Relativity, identifying $\Lambda$ with the cosmological counterterm and the physical energy-momentum tensor as $T^\mu_\nu=\lambda^4 \tilde T^\mu_\nu$. However, now we have two extra ingredients -- the two global equations (\ref{lambda}) and (\ref{Lambda}).  As we will see, these will fix the cosmological counterterm, but first note an importance consequence of equation (\ref{lambda}). Since $\sigma$ is assumed to be differentiable, if the spacetime volume is infinite, $\lambda$ is forced to vanish. However, since all the physical masses in the matter sector scale with $\lambda$ relative to the Planck scale, vanishing $\lambda$ would force them too to vanish. We do not live in a Universe in which all particle masses in the Standard Model  are zero, so we see  that the spacetime volume must be   finite. In other words, spatial sections must be finite, and the Universe must ultimately end in a crunch.

\subsubsection*{\it Cancellation of vacuum energy}
Let us now evaluate the cosmological counterterm. Combining equations (\ref{lambda}) and (\ref{Lambda}) yields 
\be
\Lambda=\frac14 \langle T^\alpha_\alpha \rangle
\ee
where the angled brackets denote the {\it spacetime} average $\langle Q \rangle=\frac{\int d^4 x \sqrt{-g} Q}{ \int d^4 x \sqrt{-g}}$. Plugging this back into equation (\ref{g}) yields 
\be \label{seqeqT}
M_{pl}^2 G^\mu_\nu = T^\mu_\nu -\frac14 \delta^\mu_\nu \langle T^\alpha_\alpha \rangle
\ee
The energy momentum tensor can be written as $T^\mu_\nu=-V_{vac} \delta^\mu_\nu+ \tau^\mu_\nu$, where $\tau^\mu_\nu$ are local excitations about the vacuum, and $V_{vac}$ is the vacuum energy coming from Standard Model loops. We can calculate the latter to any desired loop order, but it makes no difference to equation (\ref{seqeqT}). The Standard Model vacuum energy will {\it always} drop out of the gravitational dynamics, leaving us with
\be \label{seqeqtau}
M_{pl}^2 G^\mu_\nu = \tau^\mu_\nu -\frac14 \delta^\mu_\nu \langle \tau^\alpha_\alpha \rangle
\ee
This is it - there are no hidden equations like in unimodular gravity. The classical dynamics of the gravitational field is completely determined by local matter excitations, and is  independent of the vacuum energy, no matter how it is calculated.  There is  a  sense in which we have decoupled the zero modes for matter from the zero mode for gravity, reminiscent of the decapitation scenario of Adams {\it et al} \cite{decap}.

\subsubsection*{\it The importance of a universal coupling between matter and the global degrees of freedom}
The cancellation works to each and every order in loops, thanks to diffeomorphism invariance and the fact that protected matter couples universally to $\lambda$ via the metric $\tilde g_{\mu\nu}=\lambda^2 g_{\mu\nu}$. This ensures that at any order in loops, the vacuum energy scales like $\lambda^4$, which is essential for the cancellation to work at each and every order. This is in contrast to Tseytlin's model \cite{Tseytlin}, for which the tree level vacuum energy scales differently with respect to the global degrees of freedom in comparison to loop corrections\footnote{In Tseytlin's model, the tree-level vacuum energy scales as   $1/V$, where $V$ is the spacetime volume, while the loops go like $1/V^2$ \cite{KP1, KP2}.}. It is only the tree level contribution that gets cancelled in Tseytlin's model. For the sequester the cancellation is to each and every order.

\subsubsection*{\it The residual cosmological constant} 
Despite the cancellation of vacuum energy at each and every order in the sequester, we are left with a residual cosmological constant,
\be \label{Leff}
\Lambda_\text{eff}=\frac14  \langle \tau^\alpha_\alpha \rangle
\ee
but this only depends on the historic average of locally excited matter. There is no dependence on vacuum energy!  In other words, the residual cosmological constant is completely insensitive to radiative corrections that plague  zero point energies in the Standard Model. Indeed, we should think of $\Lambda_\text{eff}$ as being given by a (radiatively stable) future boundary condition. This boundary condition then fixes the {\it spacetime} average of (the trace of) the local excitations, $\langle \tau^\alpha_\alpha \rangle$. This is a global quantity characterising a particular solution.  Fixing its value does not affect {\it local} dynamics, and does not lead to any {\it local} violation of causality.  There is no obvious  pathology. On the contrary, requiring a global constraint sits well with the discussion of the previous section, and the idea that measurement of the cosmological constant can only be achieved by scanning all of space and time.

What value should we assign to $\Lambda_\text{eff}$? Well, following the philosophy outlined in section \ref{sec:ccp}, now that we are lucky enough to have a radiatively stable cosmological constant, its value should be fixed empirically by observation. This means that it should not exceed the critical density today $\rho_c \sim (\text{meV})^4$. However,  we still need to check if this empirical choice yields a  boundary condition on $\langle \tau^\alpha_\alpha \rangle$ that is compatible with  a large and old Universe like ours that has yet to undergo collapse.

Assuming homogeneity and isotropy, it turns out that for ordinary matter satisfying standard energy conditions ($-1<p/\rho \leq 1$),  the historic average $\langle \tau^\alpha_\alpha \rangle$ is dominated by its contributions near the turning point just before collapse.  In particular, we  find\footnote{For $p/\rho=1$ running into either singularity, we need to regulate a logarithmic divergence but cutting off the geometry at large curvature.} \cite{KP2}
\be
\int d^4 x \sqrt{-g}=a_\text{max}^4, \qquad \langle \tau^\alpha_\alpha \rangle \sim \rho_\text{turn}
\ee
where the characteristic maximum size of the Universe $a_{max} \sim \frac{1}{H_\text{turn}} \gtrsim \frac{1}{H_0}$ , and the characteristic energy density of matter near the turnaround $\rho_\text{turn} \sim M_{pl}^2 H^2_\text{turn} \lesssim \rho_c \sim M_{pl}^2 H_0^2 $. It follows that $\Lambda_\text{eff}$ is tiny, too small in fact to be responsible for dark energy because it is less than the current critical density. Our future boundary condition has allowed the Universe to get old and big.

 \subsubsection*{\it An approximate symmetry} The sequestering of vacuum energy is achieved thanks to two approximate symmetries, as one might have expected.  The first is an approximate scaling symmetry (becoming exact as $M_{pl} \to \infty$), 
 \be
 \lambda \to \Omega \lambda, \quad   \Lambda \to \Omega^4 \Lambda, \quad g_{\mu\nu}=\eta_{\mu\nu}+\frac{h_{\mu\nu}}{M_{pl}} \to \frac{\eta_{\mu\nu}}{\Omega^2}+\frac{h_{\mu\nu}}{M_{pl}\Omega} \quad \implies \quad S\to S+{\cal O}(1/M_{pl})
 \ee 
The second is an approximate shift symmetry (becoming exact as $\mu \to \infty$),
\be
{\cal L}_m \to {\cal L}_m+\alpha , \quad \Lambda \to \Lambda+\alpha  \lambda^4 \quad \implies \quad S\to S+{ \cal O}(\alpha/\mu^4)
\ee
These symmetries protect the curvature associated with the residual cosmological constant $\Lambda_\text{eff}/M_{pl}^2$, rendering it naturally small\footnote{$ \Lambda_\text{eff}/M_{pl}^2 \to 0$ as $M_{pl} \to \infty$ and as $\mu \to \infty$. The latter follows from the fact that if the spacetime volume is held fixed and $\mu \to \infty$, then we are in the conformal limit and the trace vanishes.} .

 \subsubsection*{\it Phase transitions don't matter}
 Now let's discuss the added complication of phase transitions.  As we will see, they will only give a negligible contribution, becoming less significant the earlier transition.  For simplicity, model the phase transition as a step function in the potential,
 \be
 V=\begin{cases} V_\text{before} & t<t_\text{PT} \\ V_\text{after} & t>t_\text{PT} \end{cases}
 \ee
 where $\Delta V =V_\text{after}-V_\text{before}<0$ is the jump in vacuum energy, and $t_\text{PT}$ is the time of the transition. The source term on the right hand side of equation (\ref{seqeqtau}) is given by
 \be
  \tau^\mu_\nu -\frac14 \delta^\mu_\nu \langle \tau^\alpha_\alpha \rangle=-V_\text{eff} \delta^\mu_\nu \ee
 where 
 \be
 V_\text{eff}=V-\langle V \rangle=\begin{cases} V^\text{eff}_\text{before} & t<t_\text{PT} \\ V^\text{eff}_\text{after} & t>t_\text{PT} \end{cases}
\ee
Through explicit calculation \cite{KP2}  we see that 
\be
V^\text{eff}_\text{after}=-|\Delta V| \epsilon_\text{PT}, \qquad V^\text{eff}_\text{before}=|\Delta V | (1-\epsilon_\text{PT})
\ee
where 
\be
\epsilon_{PT}=\frac{\int^{t_\text{PT}}_{t_\text{bang}} dt a^3}{\int^{t_\text{crunch}}_{t_\text{bang}} dt a^3} \sim \left( \frac{a_\text{PT}}{a_\text{max}}\right)^3 \frac{H_\text{turn}}{H_\text{PT}} \ll 1
\ee
Here $a_\text{PT}$ and $H_\text{PT}$ are respectively  the scale factor and the Hubble scale at the phase transition.  We expect $|\Delta V| \sim M_{pl}^2 H_\text{PT}^2$ and so
\be
|V^\text{eff}_\text{after}| \sim \rho_\text{turn}  \left( \frac{a_\text{PT}^3 H_{PT}}{a_\text{max}^3 H_\text{turn}}\right) \lesssim \rho_\text{turn} \lesssim \rho_c
\ee
where we have used the fact that standard energy conditions ($-1 <p/\rho \leq 1$) suggest that $a_\text{PT}^3 H_{PT} \lesssim a_\text{max}^3 H_\text{turn}$ \cite{KP2}. Thus the effect of the phase transition  is small at late times, becoming less significant the earlier the transition. This is because the sequestering of vacuum energy is effective after the transition, so the earlier it happens, the more cancellation. This is ideal because the QCD and electroweak phase transitions happened very early indeed, so we get plenty of suppression. We also see that  $V^\text{eff}_\text{before} \sim M_{pl}^2 H_\text{PT}^2$, suggesting that there is a short burst of inflation just before a sharp  transition like the one we have described. For smoother transitions one can obtain longer periods of inflation, so much so that the model can be made compatible with early Universe inflation, with the inflation still living in the protected matter sector \cite{KP2}.

\subsubsection*{\it The Universe must be closed}

The practical way to find sequestering solutions is as follows. We start with the following one parameter family of equations  familiar to General Relativity
\be
M_{pl}^2 G^\mu_\nu =\tau^\mu_\nu-\Lambda_\text{eff} \delta^\mu_\nu
\ee
where the ``one parameter" is of course, the residual cosmological constant, $\Lambda_\text{eff} $ which is currently left unspecified. For a given set of initial data, we then derive a one parameter family of solutions and then pick the one that satisfies the integral constraint $\langle R \rangle =0$. This fixes $\Lambda_\text{eff}$ to take on the desired form given by equation (\ref{Leff}).

Of course, for a given set of initial data, it may be that the constraint cannot be satisfied in which case there is no corresponding sequestering solution. Or it may be that a solution exists, but does not lead to a large and old Universe, in which case that is not a sequestering solution that can describe our Universe.  Generically, however, we find that suitable sequestering solutions do exist, although in cosmology they will be constrained to be spatially closed.

To see this,  consider Friedmann-Robertson-Walker solutions with spatial curvature $k$, in the presence of a fluid with equations of state, $w$, and an (as yet)  unspecified residual cosmological constant, $\Lambda_\text{eff}$. The family of Friedmann equations is given by
\be
3 M_{pl}^2 \left(H^2 +\frac{k}{a^2}\right) = \frac{\rho_0}{a^{3(1+w)}}+\Lambda_\text{eff}
\ee
for some constant $\rho_0>0$. This can be conveniently written as $k=U(a)-\dot a^2$, where $U(a)=\frac{\Lambda_\text{eff}} {3 M_{pl}^2} a^2+\frac{\rho_0}{3 M_{pl}^2} a^{-1-3w}$, from which we infer that real solutions can only exist provided $k \leq U(a)$.

In GR,  for any given value of $k$ real solutions will always exist for a range of $\Lambda_\text{eff}$.  This is not true in sequestering because the integral constraint $\langle R \rangle =0$ gives
\be
\int dt a^3 \left[\dot H+2H^2+\frac{k}{a^2} \right]=0 \implies k=\frac{-[a^3 H]^\text{crunch}_\text{bang}+\int dt a^3 H^2}{\int dt \frac{1}{a}}
\ee
Given that $H>0$ at the bang, and $H<0$ at the crunch, we see that $[a^3 H]^\text{crunch}_\text{bang}<0$, and so we conclude $k>0$. Strictly speaking the limits here (labelled bang and crunch) represent some cut-off  at high curvature and one ought, in principle, to revisit this argument allowing for possible corrections beyond the cut-off. However, in a large and old Universe like our own, such corrections should not be significant and the requirement of $k>0$ remains valid. For Universes that never get beyond tiny, the UV corrections may prove important and one should probably relax the bound on $k$.

\subsubsection*{\it Transient dark energy }
A dark energy equation of state $w=-1$ would lead to an eternal Universe, expanding forever at an accelerated rate. Such a scenario is incompatible with the sequestering proposal, since it would lead to an infinite spacetime volume, and therefore all particle masses would have to vanish. It follows that dark energy must be transient in a sequestering Universe, with $w \approx -1$ being merely an approximation. 
\subsubsection*{\it The End}
There is very simple dark energy model that when embedded within sequestering yields a possible explanation as to why dark energy has just begun to dominate \cite{KP3}.  To see this, we first note that a realistic sequestering Universe must have two vital ingredients: it must contain dark energy, and it must ultimately collapse. Being as conservative as possible, let us suppose that the same field is responsible for both. We will also require that the potentials governing the dynamics of this field are technically natural and do not suffer from any radiative instabilities. 

There is a dark energy theory that satisfies all of the above: the linear potential. This is one of the first ever models of quintessence\cite{andrei87} in which a canonical  scalar field, $\phi$,  minimally coupled to gravity moves under the influence of a potential
\be \label{lin}
V=m^3 \phi
\ee
We shall embed this in the sequestering scenario by including the scalar in the protected matter sector. The potential (\ref{lin}) is radiatively stable, its form protected by a shift symmetry\footnote{When embedded in GR rather than sequestering, this is reduced to a {\it pseudo-symmetry}.} 
$$\phi \to \phi+c, \qquad \Lambda \to \Lambda+m^3 c\lambda^4$$
The size of $m^3$ is also technically natural, as radiative corrections to it always involve graviton loops, and are therefore Planck suppressed. 

Once the scalar begins to dominate, it does so in {\it slow roll,} provided the initial value of $\phi$ during this epoc, $\phi_\text{in} \gtrsim M_{pl}$.  Slow roll yields a period of cosmic acceleration until cosmological collapse
occurs at a time \cite{KP3}
\be
t_\text{collapse} \sim \sqrt{\frac{M_{pl}}{m^3}}
\ee
Collapse is guaranteed as the scalar rolls to sufficiently negative values, the potential providing the required negative energy.  Furthermore, the time of collapse is radiatively stable, through its dependence on the radiatively stable parameter $m^3$. We can therefore choose it in accordance with observation, confident in the knowledge that such a choice is invulnerable to loop corrections. In order to postpone collapse until at least the current epic we require $m^3 \lesssim M_{pl} H_0^2$.

 Is slow roll guaranteed? When the linear potential is embedded in GR, the answer is ``no". One has to fine tune the initial value $\phi_\text{in}$ such that it is super-Planckian. And this really is a fine tuning in the sense that $\phi_\text{in}$ is {\it not} a radiatively stable quantity in GR. This is because the radiative instability of $\Lambda$ is transferred to $\phi_\text{in}$ once the the pseudo-symmetry is used to fix the former. The situation for sequestering is much better, and slow roll {\it is} guaranteed without any fine tuning. This is because the dynamical sequestering constraint $\langle R \rangle=0$ happens to select precisely those GR solutions that have $\phi_\text{in} \gtrsim M_{pl}$.  This is remarkable. Sequestering {\it guarantees} acceleration just before collapse in complete constrast to GR!
 
 Returning to the question of  dark energy and {\it why now}, our answer should be {\it because the end is nigh}. And why is it nigh? Because the radiatively stable parameter $m^3 \sim M_{pl}H_0^2$. Existing and future surveys (eg. DES, Euclid) should help to constrain this model, and some initial work is already under way \cite{jon}

\subsubsection*{\it Sequester, where art thou?}
At this stage we view the sequester as an effective theory, valid in the semi-classical limit in which quantum gravity effects are ignored, but matter loops are important.  This is precisely the regime in which the cosmological constant problem is most sharply formulated, and we have seen how it may be alleviated. Having said that, it is natural to ask how the sequester might arise in fundamental theory.  To this end  it is illuminating to change frames to ``Jordan frame", by redefining our variables $(g_{\mu\nu}, \Lambda, \lambda) \to (\tilde g_{\mu\nu}, \tilde \Lambda, \tilde M_{pl})$ where 
\be
\tilde g_{\mu\nu}=\lambda^2 g_{\mu\nu}, \qquad \tilde M_{pl}=\frac{M_{pl}}{\lambda}, \qquad \tilde \Lambda=\frac{\Lambda}{\lambda^4}
\ee
The Jordan frame action now reads
\be \label{seqactj}
S=\int d^4 x \sqrt{-\tilde g}\left[ \frac{\tilde M_{pl}^2 }{2} \tilde R-{\cal L}_m (\tilde g^{\mu\nu}, \Psi)-\tilde \Lambda \right]+\sigma \left(\frac{\tilde \Lambda}{\mu^4} \right)
\ee
so that  the variation with respect to $\tilde M_{pl}$ now yields the global constraint $\int d^4 x \sqrt{-\tilde g} \tilde R=0$, which is key to the cancellation mechanism. The form of the action (\ref{seqactj}) is consistent with promoting fundamental constants to dynamical variables. Allowing fundamental constants to vary in this way is reminiscent of string compactifications \cite{comp} or wormhole corrections in Euclidean quantum gravity \cite{cole}. Indeed, one might hope that a sum over compactifications or wormhole topologies (see section \ref{sec:cole}) might yield a sequestering path integral schematically of the form
\be
\int  d M_{pl} \int d \Lambda \int D g_{\mu\nu} \int D \Psi e^{i S[g, \Psi; M_{pl}, \Lambda)}
\ee
where the integration over $M_{pl}$ and $\Lambda$ is ordinary high school integration, rather than functional integration. In such a scenario, the sequestering function $\sigma$ would presumably come from a non-trivial measure in the sum.

\section{Some final thoughts}  \label{sec:summary}
Back in the 1930s physics was thrown into crisis by the divergent loop corrections to quantum electrodynamics. Did this demand a {\it conservatively radical} approach in which we abandoned QED and sought after a new theory? Or were we to take a {\it radically conservative}\footnote{This terminology is originally due to Wheeler.}   perspective, and  endeavour to better understand QED itself?  As we now know, the radical conservatives  won through as we built up our understanding of quantum field theory and renormalisation.  The cosmological constant problem smells of a similar kind of crisis. Does it call for radical conservativism or conservative radicalism?  The  robustness of GR as the only consistent  low energy theory for interacting massless spin $2$ particles could be taken as a warning against too much radicalism (see e.g. \cite{nima,nima2}). Is it really OK to modify gravity when faced with the cosmological constant problem?  Weinberg's no go theorem \cite{Wein} makes the case against even stronger.

In these lecture notes we have tried to review the state of play, highlighting the most interesting approaches to the cosmological constant problem.  Being radically conservative, we could keep faith with GR at low energies, and seek a resolution through symmetry, but none of the symmetries discussed in section \ref{sec:symm} are quite up to the job, at least not in our Universe.  Could some UV modification of gravity help? As we discussed in section \ref{sec:short} the challenge then is to connect ultrasoft gravitons probing cosmology at the horizon scale to new gravitational physics at a millimeter or below. Indeed, when we have an ultrasoft external graviton connected to a matter loop, the high momentum contribution from the loop must be cancelled or suppressed, but only when there are no external Standard Model fields.  If external Standard Model fields are present, we know from our discussion of section \ref{sec:exist} that the high momentum contributions to the  loop do in fact gravitate.

And so in desperation we are faced with two alternatives: the multiverse and anthropic selection, or an infra-red modification of gravity. In the latter, there be dragons: low cut-offs, ghosts, acausality, all of which reinforce the  robustness of General Relativity. However, there is one way to modify gravity in the infra-red whilst respecting all the usual consistency arguments: a {\it global} modification. This is the sequester \cite{KP1,KP2}. There is a definite sense in which the sequester falls into the category of radical conservatism. Locally this proposal is indistinguishable from General Relativity, so it happily respects the inevitability of GR from a consistency perspective. The deviation is only at a global level, and yet this cuts straight to the core of the cosmological constant, which we can think of as the global component of the stress energy tensor. This allows one to sequester from gravity all radiative corrections  to the vacuum energy from a protected matter sector, which we take to include the Standard Model. The result is a residual cosmological constant that is radiatively stable, insensitive to matter loop corrections to the vacuum energy at any order in perturbation theory. 

The sequester does not {\it predict} the value of the residual cosmological constant, but nor should it.  Like the coupling of any operator of dimension $\leq 4$, it value cannot be predicted because of the need to regulate divergences, and introduce an arbitrary subtraction scale. But thats just renormalisation. You do not predict these couplings, you {\it measure} them. But you want to be sure that the measurement is robust against further radiative corrections. In some respects, we can think of the cosmological constant as a yardstick for cosmology, which we then use to calibrate other cosmological sources through their relative effect on curvature.  You can imagine something similar in Newtonian physics, a single inertial mass that calibrates other inertial masses. In fact without this Newtonian yardstick, Newton's second law is rendered virtually vacuous, as would be the Friedmann equation without our cosmological yardstick.  In Newtonian physics, if we double the force and measure the same acceleration, we conclude that the mass has doubled, but  we cannot actually {\it predict} the mass of any one object. To do that we must first  set up our yardstick by comparing the force versus acceleration for our single inertial mass, and from there all others may be calibrated. It  follows that the yardstick itself is not something you should ever hope to predict. But you do want to be able to trust its stability.  What good is a yardstick that changes its size drastically whenever you alter your effective description? You need a {\it radiatively stable} yardstick. This is the status of the cosmological constant in sequestering: a radiatively stable yardstick for cosmology.

The sequester  represents a new approach to the cosmological constant problem that embraces both General Relativity and renormalisation. So in a sense, there is no radical new physics, just a radical new way of understanding the physics. The proposal is  still in the early stages of development,  and there is much to learn,  but at least the radical conservative now has something more than the landscape to consider.

\section*{Acknowledgements}
I am indebted to my long term collaborator, Nemanja Kaloper, whose insight and understanding is behind much of these lecture notes. Thanks also to David Stefanyszyn for interesting discussions and carefully reading through this manuscript.  I would also like to thank the organisers of the Mexican school for their wonderful  hospitality, and especially Gustavo Niz, Aileen and Matias for providing excellent company. AP was funded by a Royal Society URF.

\end{document}